\documentclass[nofootinbib,aps,prd,reprint, preprintnumbers]{revtex4-2}
\usepackage[utf8]{inputenc}
\date{2024}

\usepackage{eurosym}
\usepackage{graphicx}
\usepackage{amsmath}
\usepackage{mathrsfs}
\usepackage{bm}
\usepackage{color}
\usepackage{amsmath}
\usepackage{amssymb}
\usepackage{float}
\usepackage{multirow}
\usepackage{array}
\usepackage{subfig}
\usepackage{leftindex}
\usepackage{mleftright}
\usepackage{xparse}

\newcommand{\fixme}[1]{\textcolor{red}}

\allowdisplaybreaks

\begin{document}
    
\title{Q-ball collisions and their Gravitational Waves}
\author{Deog Ki Hong}
\email{dkhong@pusan.ac.kr}
\author{Stephen J. Lonsdale}
\email{stephenjlonsdale@gmail.com}
\affiliation{Department of Physics, Pusan National University, Busan 46241, Korea\\
Extreme Physics Institute, Pusan National University, Busan 46241, Korea}

\date{\today}

\begin{abstract}
    
Sydney Coleman's Q-ball remains a compelling instance of the formation of localised objects within classical field theory, independently of the quantum evolution. The theoretical possibility of such objects forming and colliding in the early universe from models such as Affleck-Dine fragmentation, or from a number of mechanisms where they are produced copiously with various sizes and charges to constitute dark matter in the universe, makes it important to study in detail Q-ball collision phenomenology. In this work we present certain results from the study of Q-ball collisions and their generation of gravitational waves, using a new general code package for scalar fields coupled to gravity. We then comment on the possibility of future gravitational wave detectors searching for signals of Q-ball collisions.
    
\vspace{5mm}
    
    
\end{abstract}

\maketitle
\preprint{PNUTP-24/A04}

\section{Introduction }

Coleman's Q-ball provides a clear example of solitons in quantum field theory, displaying a capacity for the formation of compact objects via classical dynamics \cite{COLEMAN1985263,Rosen:1968mfz}.
In the early dynamics of the universe, such compact objects may play a pivotal role in establishing the conditions for dark matter production or baryogenesis.  Alternatively solitons exist as a dark matter candidate in their own right. Collisions between large Q-balls that conserve Noether charge may dissipate energy via gravitational waves, making the phenomenology of Q-ball collisions at large scale an interesting area of study.
This may also be relevant to the future of gravitational wave astronomy where we may either see signals of such objects generated in the early universe or detect their collision events. 
\\\\

 Being non-topological solitons, Q-balls are energetically stable by the conservation of a Noether charge. Their stability against decay is then guaranteed by having an energy per unit charge that is  less than an equivalent number of charged quanta.
This stability condition also depends on the specific scalar potentials that feature attractive terms beyond the quadratic terms, which may exist in a number of theories beyond the standard model (BSM).
With the stability condition satisfied, Q-balls are stable against fission into smaller Q-balls and may preferentially fuse in the case of collisions.  Refs.~\cite{marieke,Croon_2020} demonstrates models that bridge the gap between first-order phase transitions and Q-balls in solitosynthesis. The excess energy resulting from the fusion of Q-balls may lead to excited states or be dissipated by the emission of uncharged fields if coupled. 
\\\\

We follow the previous numerical and analytical study of Q-ball profiles done in~\cite{Ioannidou:2003ev,Battye_2000, Volkov_2002}, together with their application to the phase transition in condensed matter physics \cite{Hong:1987ur}, and of particular interest, cosmological applications. In many models of early universe dynamics, complex scalar fields may attain large vacuum expectation values (VEVs) and non-zero angular velocity in their phase, creating Affleck-Dine (AD) condensates. Supersymmetric models in particular are expected to be capable of forming AD condensates due to the requirements of flat directions in potentials. For example the minimal supersymmetric standard model (MSSM) has hundreds of flat directions~\cite{Gherghetta_1996}. The fate of AD condensates following the fragmentation of the field, being itself a source of gravitational waves~\cite{Kusenko_2008,Tsumagari:2009na,Chiba_2010,Zhou_2015,white:2021hwi,Hou:2022jcd,Lozanov_2024}, can result in Q-balls which may be long or short lived depending on the model, and these can be generated with large velocities in a period of possibly frequent collisions \cite{Chiba_2010}.
\\\\

The relative velocity of colliding Q-balls has been shown to be the key factor in collisions to determine whether they merge, pass through each other or undergo certain fragmentation processes into a set of new dispersed smaller Q-balls~\cite{Multam_ki_2000}. During this period we may also observe radially excited Q-balls with spherical symmetry or excited Q-balls with time-dependent non-spherical profiles where energy is stored in distinct oscillations associated with the scalar modulus of the complex field. Q-balls that can dissipate energy may settle down to some combination of stable solutions. In the absence of couplings to other fields, the energy loss is expected to be split between gravitational waves and smaller Q-balls. 
\\

Past studies of head-on collisions of Boson-stars in \cite{Palenzuela_2007,Atteneder_2024,Evstafyeva_2023,Bezares:2017mzk,Ruiz_2007} discuss the quadrupolar oscillation that results from the merged states of stars carrying Noether charges. Recent work has also looked at gauged-Q-ball collisions \cite{Anagnostopoulos:2001,PhysRevD.110.015012, kinach2024dynamicsu1gaugedqballs}. In Ref. \cite{Battye_2000} ring structures emerging from head-on Q-ball collisions are found for particular cases of high relative-velocity and various relative phases in 3D simulations. Q-rings are studied analytically in \cite{Axenides:2001jj} to discuss such objects and their lifetime.  The collision phenomenology of 2D Q-balls is also examined in detail in~\cite{Axenides:1999hs}.
\\\\

With growing size and charge,  the relevant self-gravitational effects in the Q-ball collision require proper treatment as in the case of Boson-stars.   
The limit for how large a Q-ball can be has been studied for example in~\cite{Sakai:2011wn}.
The collision dynamics of Q-balls from scalars with couplings to gravity may be found by considering the classical evolution of combined scalar fields with the Einstein-Hilbert (E.H.) action in a lattice simulation where all scalar fields are evolved by symplectic integrator methods. In this work we construct such a general symplectic integrator for $n$ scalars coupled to gravity. Following the work of HLATTICE, we develop an original python code for Q-Ball collisions in a weak field regime of general relativity (GR).
\\\\

The structure of the work is as follows. Section II is an introduction to Q-ball physics. In Section III we review the numerical method for simulating collisions coupled to gravity. In Section IV we present a number of highly boosted collision simulation cases including excited-state formation, forward scattering, Q-ball/anti-Q-ball head-on collisions and a Q-ring formation collision. Section V concludes with a discussion of future cases of interest. 

\section{Q-Ball Field Profiles} 

For a single complex scalar field with U(1) global symmetry, arising from U(1) symmetric potential $V(\phi)$, the charge of scalar fields is given as 
\begin{equation}
Q=\frac{1}{2i} \int d^3x (\phi^* \partial_t \phi - \phi \partial_t \phi^*).
\end{equation}
Similarly, the energy within the field is defined as 
\begin{equation}
E= \int d^3 x \left[\frac{1}{2} |\dot{\phi}|^2 +\frac{1}{2} |\nabla \phi|^2 + V(\phi)\right].
\end{equation}

Dictated by the time-translation invariance of the system,
the field carrying a fixed charge takes the form $\phi= e^{i \omega t } \phi(x)$ and its charge is given by 
\begin{equation}
Q= \omega \int \left|\phi(x)\right|^2 d^3x.
\end{equation}
The Q-ball solution for fixed charge can be found via numerical shooting method for the effective equation of motion in D space dimensions:
\begin{equation}
\frac{d^2 \phi}{dr^2} +\frac{(D-1)}{r}\frac{d \phi}{dr}=  V'(\phi) - \omega^2\,\phi.
\end{equation}
The thin-wall ground state solutions typically have a constant field profile out to radius R with fixed angular frequency, while thick-wall solutions for smaller charge and larger $\omega$ interpolate over the full radius of the spherically symmetric state. The boundary conditions $\phi'(0)=0,\, \phi(\infty)=0$ are imposed and the initial field value varied until a stable solution is found.
\begin{figure}[h]
\centering 
\includegraphics[width=0.4\textwidth]{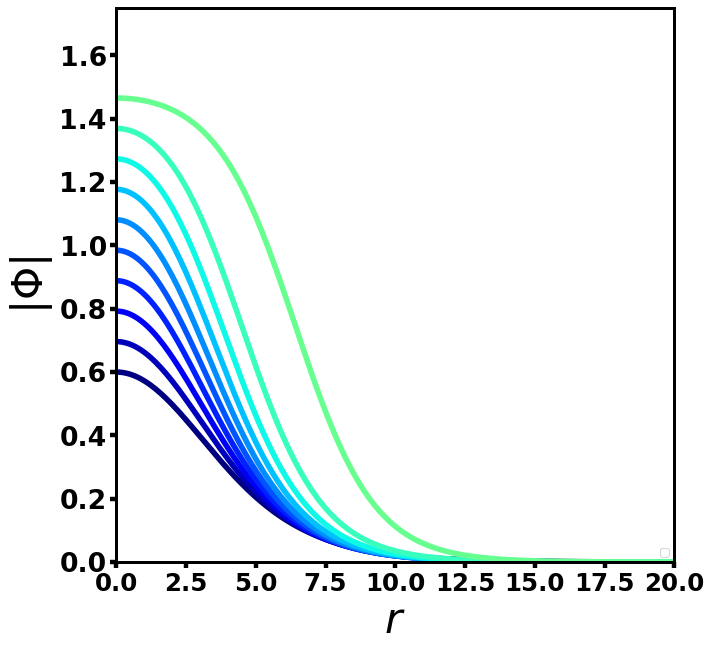}
\label{profile}
\caption{Sample profiles of Q-balls for $\omega^2=0.8^2 -0.98^2$ from the shooting method. Throughout this work we will use these sample Q-ball profiles with $A=1,B=0.5, M=1$ in potential $V_I$, introduced in subsection~A. (All the dimensional quantities are in units of $M$ throughout the paper.)   }
\end{figure}

\subsection{Complex Scalar Potentials}

We can examine a number of different class of   potentials for complex scalars, motivated by different BSM models. Among them we take the simplest polynomial potential, given as 
\begin{equation}
V_{I}= M^2 |\phi|^2 -A |\phi|^3 +B |\phi|^4.
\end{equation}
Any suitable potential that admits Q-ball solutions can be used, though, to find an initial profile to set up collision events.  A range of profiles for different angular frequencies is plotted in Fig.~1, using the shooting method, that forms initial states for collision events. These profiles are spherically symmetric in their rest frame and so only the radial profiles are shown. 

\subsection{Stability}
The total energy of a Q-ball should be less than $mQ$, the energy of an equivalent number of scalar quanta of mass $m$. For spherically symmetric profile, the solution takes the following form, taking the center of Q-ball to be the origin, $r=0$,  
\begin{equation}
\phi(x,t)= \sigma(r) e^{-i\omega t }.
\end{equation}
For large Q so that the thin-wall approximation is good,  the radius scaling comes from the definition of charge with step function at the boundary,
\begin{equation}
Q=\frac{4}{3} \pi R^3 \omega\phi_0^2,
\end{equation}
where $\phi_0=\sigma(0)$.
The `ground state' solutions are equivalently classified by their magnitude at the origin, radius, charge and angular phase frequency for a given choice of potential. The relative velocity and relative phase of Q-balls are then additional parameters for the collisions.

\section{Numerical Methods}
For $n$ scalar fields $\phi_i$ ($i=1,\cdots,n$), we have
\begin{equation}
\mathcal{S}=\int d^4 x\left(\frac{1}{2}\sum_{i=1}^n\partial^\mu \phi_i\partial_\mu \phi_{i} -V(\phi_1,...,\phi_n) +\frac{M_p^2}{2}R \right).
\label{EH}
\end{equation}
In order to simulate Q-ball collisions and dynamics along with gravitational wave signals, we make use of the symplectic integrator method and symplectic factorization used in Hlattice 2.0 \cite{Huang_2011} where it is described in detail. The present work uses the original python code to implement this method.  This retains the use of synchronous gauge. Additionally, all of the variables are defined in terms of fields and their conjugate momenta for the scalar sector, and metric perturbations with their effective conjugate momenta for the gravitational variables. Components of a state vector $f$, which is defined at each point of an $L^3$ lattice, are updated by time evolution operator for discrete time jumps of size $dt$,
\begin{equation}
f(t+dt)= e^{{\bold H} dt} f(t). 
\end{equation}
Dividing the H operator into a sum of possibly non-commuting components we can write a symplectic integrator factorisation,
that is, for arbitrary operators A,B,C, the factorisation, 
\begin{align}
e^{(A+B+C)dt}&=e^{c_3 A dt/2}e^{c_3 B dt/2} e^{c_3 C dt/2} e^{c_3 B dt/2} e^{(c_3+c_2) A dt/2}  \nonumber \\
&\times e^{c_2 B dt/2} e^{c_2 C dt/2} e^{c_2 B dt/2} e^{(c_2+c_1) A dt/2} \nonumber\\
&\times e^{c_1 B dt/2} e^{c_1 C dt/2} e^{c_1 B dt/2} e^{(c_1+c_0) A dt/2} \nonumber\\
&\times e^{c_0 B dt/2} e^{c_0 C dt/2} e^{c_0 B dt/2} e^{(c_0+c_1) A dt/2} \\
&\times e^{c_1 B dt/2} e^{c_1 C dt/2} e^{c_1 B dt/2} e^{(c_1+c_2) A dt/2} \nonumber\\
&\times e^{c_2 B dt/2} e^{c_1 C dt/2} e^{c_1 B dt/2} e^{(c_1+c_2) A dt/2} \nonumber\\
&\times e^{c_3 B dt/2} e^{c_1 C dt/2} e^{c_1 B dt/2} e^{(c_1+c_2) A dt/2} +\mathcal{O}(dt^7) \nonumber
\end{align}
where the coefficients \cite{Yoshida:1990zz} are
\begin{align}
c_1&=-1.17767998417887, \nonumber \\
c_2&=0.235573213359357,\\
c_3&=0.784513610477560,\nonumber \\
c_0&=1-2(c_1 +c_2 +c_3)\nonumber 
\end{align}
and is accurate to the sixth order in $dt$ as long as we can express each of the individual operators $e^A,e^B,e^C$.

Depending on the evolution of the scale factor, we may treat the lattice setup and the collision in a Friedmann–Lemaître–Robertson–Walker (FLRW) metric or in Minkowski space.
In the weak field limit the last term in the action, Eq.~(\ref{EH}), can be treated as, following~\cite{Huang_2011},
\begin{align}
&\mathcal{S_R}=\int dt \left[ -\frac{M_p^2}{4}\int g^{\frac{1}{2}} d^3 x (\dot{\beta}_{23}^2 +\dot{\beta}_{31}^2+\dot{\beta}_{12}^2\right.\nonumber \\ 
&-\dot{\beta}_{11}\dot{\beta}_{22}- \dot{\beta}_{22}\dot{\beta}_{33} - \dot{\beta}_{33}\dot{\beta}_{11}  ) \nonumber \\ 
&- \frac{M_p^2}{4} a(t)\int  d^3 x (\beta_{23,1}^2 +\beta_{31,2}^2 +\beta_{12,3}^2 -2 \beta_{23,1}\beta_{31,2}\nonumber \\
 &-2 \beta_{31,2}\beta_{12,3}-2 \beta_{12,3}\beta_{23,1}
 -\beta_{22,1}\beta_{33,1} \\
&-\beta_{33,2}\beta_{11,2} -\beta_{11,3}\beta_{22,3}\nonumber\\
&+2\beta_{23,2}\beta_{11,3}+2\beta_{31,3}\beta_{22,1}
+2\beta_{12,1}\beta_{33,2}
)  \Big{]}\,\nonumber ,
\end{align}
where $g_{ij}=a(t)^2 (\delta_{ij}+h_{ij})$,  $\beta_{ij}= ln(g_{ij})$, $\beta={\text{Tr}}(\beta_{ij})$,  $\beta_{ij,k}=\partial_k\beta_{ij}$ and at the linear order 
\begin{equation}
h_{ij}\sim \beta_{ij}- 2\delta_{ij}ln(a)\,,
\end{equation}
ignoring the gravity self-interactions. The numerical expressions for the symplectic Hamiltonian can be expressed as by dividing the Hamiltonian separately into kinetic ($K$), gradient and potential terms ($P$),
\begin{equation}
H=K+P\,.
\end{equation}
The kinetic term is further divided into diagonal and off-diagonal terms, $K=K_1+K_2$, with conjugate momenta defined as $\Pi_n=g^{\frac{1}{2}}\dot{\phi}_n$, $\Pi_{\beta_{ii}}=\frac{M_p^2}{4}g^{\frac{1}{2}}(\dot{\beta}_{ii}-\dot{\beta})$, $\Pi_{\beta_{ij}}=\frac{M_p^2}{2}g^{\frac{1}{2}}\dot{\beta}_{ij}$, where 
\begin{equation}
K_1= \sum_{\text{lat}} g^{-\frac{1}{2}} \left[ \frac{\Pi_n^2}{2} +\frac{1}{M_{p}^2}(\Pi_{\beta_{23}}^2 +\Pi_{\beta_{12}}^2+\Pi_{\beta_{13}}^2  ) \right]
\end{equation}
and
\begin{equation}
K_2= \frac{1}{M_{p}^2} \sum_{\text{lat}} g^{-\frac{1}{2}} \left[ 2 \sum_i^3 \Pi_{\beta_{ii}}^2 -(\sum_i^3 \Pi_{\beta_{ii}})^2 \right],
\end{equation}
while the sum of all gradient and potential terms becomes
\begin{align}
P = \sum_{\text{lat}} g^\frac{1}{2}\left[V +\frac{1}{2}g^{ij}\partial_i \phi \partial_j \phi\right] + \frac{M_p^2}{4n} \left(\sum_{\text{lat}} g^{\frac{1}{2}}\right)^{1/3}
\nonumber \\ \times\sum_{\text{lat}} \Big{(}\beta_{23,1}^2 +\beta_{31,2}^2 +\beta_{12,3}^2 \nonumber\\-2 \beta_{23,1}\beta_{31,2}
 -2 \beta_{31,2}\beta_{12,3}-2 \beta_{12,3}\beta_{23,1}\nonumber\\
 -\beta_{22,1}\beta_{33,1}
-\beta_{33,2}\beta_{11,2} -\beta_{11,3}\beta_{22,3}\nonumber\\
+2\beta_{23,2}\beta_{11,3}+2\beta_{31,3}\beta_{22,1}
+2\beta_{12,1}\beta_{33,2}
\Big{)}.
\end{align}

The operators act on state vector at each lattice site for time interval $dt$ as for a Hamiltonian of free particles, and with the RHS evaluated at each lattice site,  e.g.

\begin{equation}
e^{K_1 dt}     \begin{bmatrix}
           \phi_1 \\           
           \vdots\\
           \Pi_1\\
           \beta_{11}\\
           \beta_{22}\\
           \beta_{33}\\
           \beta_{23}\\
           \beta_{31}\\
            \beta_{12}\\
           \Pi_{\beta_{11}}\\
           \Pi_{\beta_{22}}\\
           \Pi_{\beta_{33}}\\
            \Pi_{\beta_{23}}\\
             \Pi_{\beta_{31}}\\
              \Pi_{\beta_{12}}
          \end{bmatrix} 
     =
          \begin{bmatrix}
           \phi_1 +g^{-1/2}\Pi_1 dt  \\           
           \vdots\\
           \Pi_1\\
           \beta_{11}\\
           \beta_{22}\\
           \beta_{33}\\
           \beta_{23} + \frac{2 g^{-1/2}}{M_p^2} \Pi_{\beta_{23}}dt \\
           \beta_{31}+ \frac{2 g^{-1/2}}{M_p^2} \Pi_{\beta_{31}}dt\\
            \beta_{12}+ \frac{2 g^{-1/2}}{M_p^2} \Pi_{\beta_{12}}dt\\\\\Pi_{\beta_{11}} +\frac{K_1}{2}dt\\
           \Pi_{\beta_{22}} +\frac{K_1}{2}dt\\
           \Pi_{\beta_{33}} +\frac{K_1}{2}dt\\
            \Pi_{\beta_{23}}\\
             \Pi_{\beta_{31}}\\
              \Pi_{\beta_{12}}
          \end{bmatrix} 
\end{equation}

\begin{equation}
e^{K_2 dt}     \begin{bmatrix}
           \phi_1 \\           
           \vdots\\
           \Pi_1\\
           \beta_{11}\\
           \beta_{22}\\
           \beta_{33}\\
           \beta_{23}\\
           \beta_{31}\\
            \beta_{12}\\
           \Pi_{\beta_{11}}\\
           \Pi_{\beta_{22}}\\
           \Pi_{\beta_{33}}\\
            \Pi_{\beta_{23}}\\
             \Pi_{\beta_{31}}\\
              \Pi_{\beta_{12}}
          \end{bmatrix} 
     =
          \begin{bmatrix}
           \phi_1 +g^{-1/2}\Pi_1 dt  \\           
           \vdots\\
           \Pi_1\\
           \beta_{11} +\frac{2g^{-\frac{1}{2}}}{M_p^2}(\Pi_{\beta_{11}}-\Pi_{\beta_{22}}-\Pi_{\beta_{33}} )dt\\
           \beta_{22}+\frac{2g^{-\frac{1}{2}}}{M_p^2}(\Pi_{\beta_{22}}-\Pi_{\beta_{33}}-\Pi_{\beta_{11}} )dt\\
           \beta_{33}+\frac{2g^{-\frac{1}{2}}}{M_p^2}(\Pi_{\beta_{33}}-\Pi_{\beta_{11}}-\Pi_{\beta_{22}} )dt\\
           \beta_{23} \\
           \beta_{31}\\
            \beta_{12} \\
            \Pi_{\beta_{11}} +\frac{K_2}{2}dt\\
           \Pi_{\beta_{22}} +\frac{K_2}{2}dt\\
           \Pi_{\beta_{33}} +\frac{K_2}{2}dt\\
            \Pi_{\beta_{23}}\\
             \Pi_{\beta_{31}}\\
              \Pi_{\beta_{12}}
          \end{bmatrix} 
\end{equation}
\\
with a lattice site label $|_{i,j,k}$ omitted on each side. Note that as in Ref. \cite{Huang_2011} we use a fourth-order Runge-Kutta for the $K_2$ operator. Symplectic integrator methods are often used for long term evolution of many body systems owing to their long term stability. In this case we are interested in the long term conservation of the global U(1) charge during the simulation and conservation of a general form of energy in the emission from gravitational waves and other light scalars. Over the course of each simulation we track the conservation of charge Q and all scalar field energy, as well as effective gravitational field energy. In long-running simulations the errors can become sizable and so long term dissipation of energy of excited states via gravitational waves is not feasible, we are however able to examine a few radial oscillation cycles to examine the dissipated gravitational wave energy in this time frame.

\begin{figure}[h]
\centering
\includegraphics[width=0.5\textwidth]{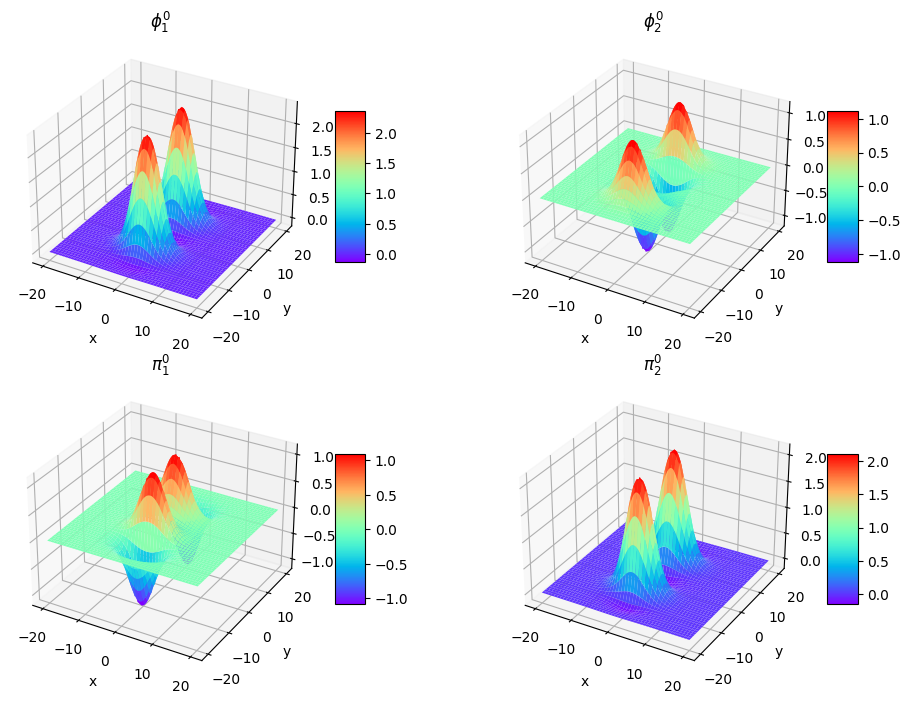}
\caption{Initial setup of two boosted Q-balls arranged to collide at the centre of the lattice grid. The four images are the Cartesian representation of the complex field and associated conjugate momenta fields.}
\end{figure}

\begin{figure*}[t]
\centering
\includegraphics[width=0.78\textwidth]{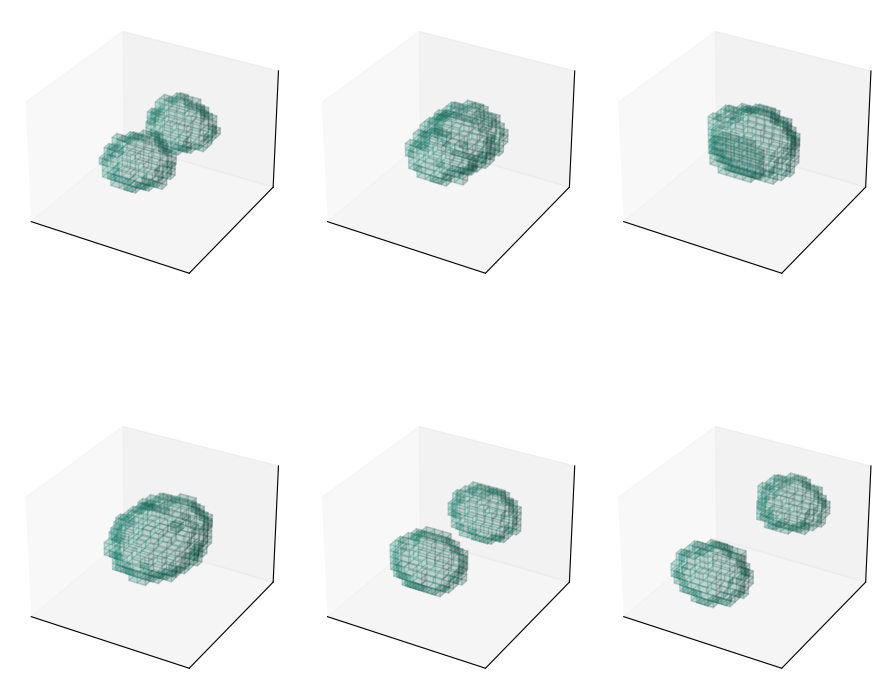}
\caption{Complex scalar energy density of two equal charge Q-balls in head-on collision. Above the critical velocity $v_C$, here at $v=0.6c$, Q-balls of equal charge undergo forward scattering with minimal interactions. The Q-ball for these cases is $\omega^2=0.806^2$. }
\end{figure*}

\begin{figure*}[t]
\centering
\includegraphics[width=.33\textwidth]{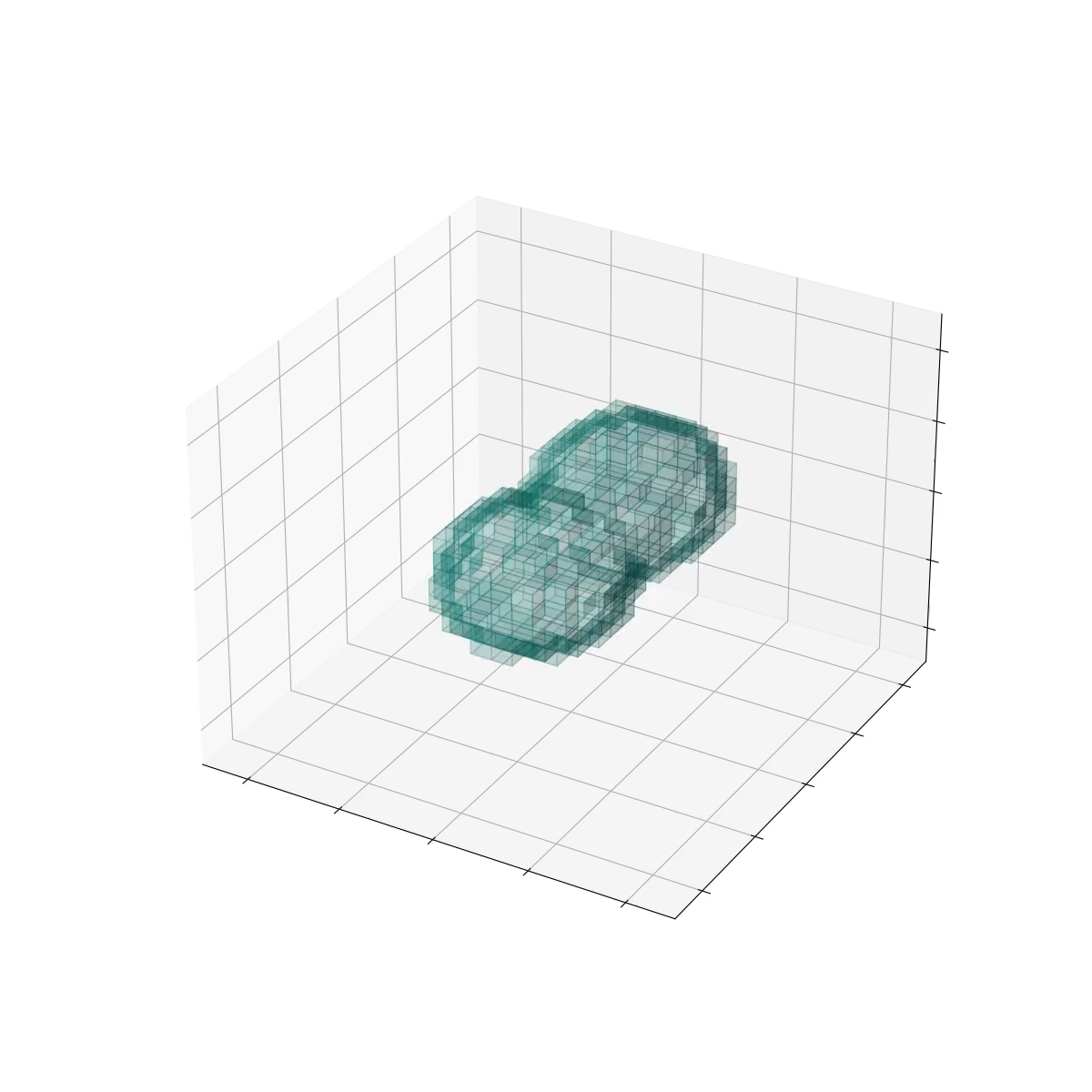}\hfill
\includegraphics[width=.33\textwidth]{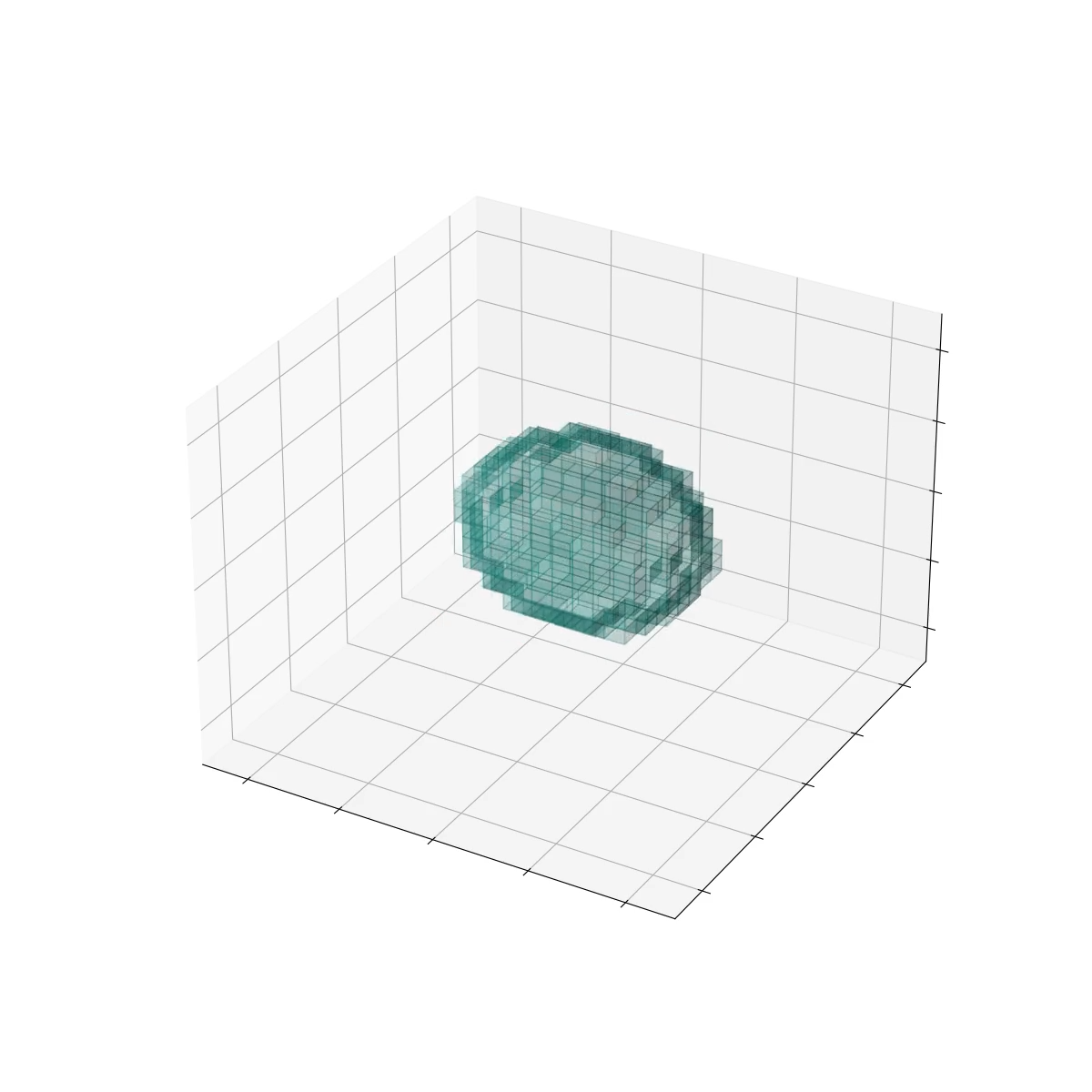}\hfill
\includegraphics[width=.33\textwidth]{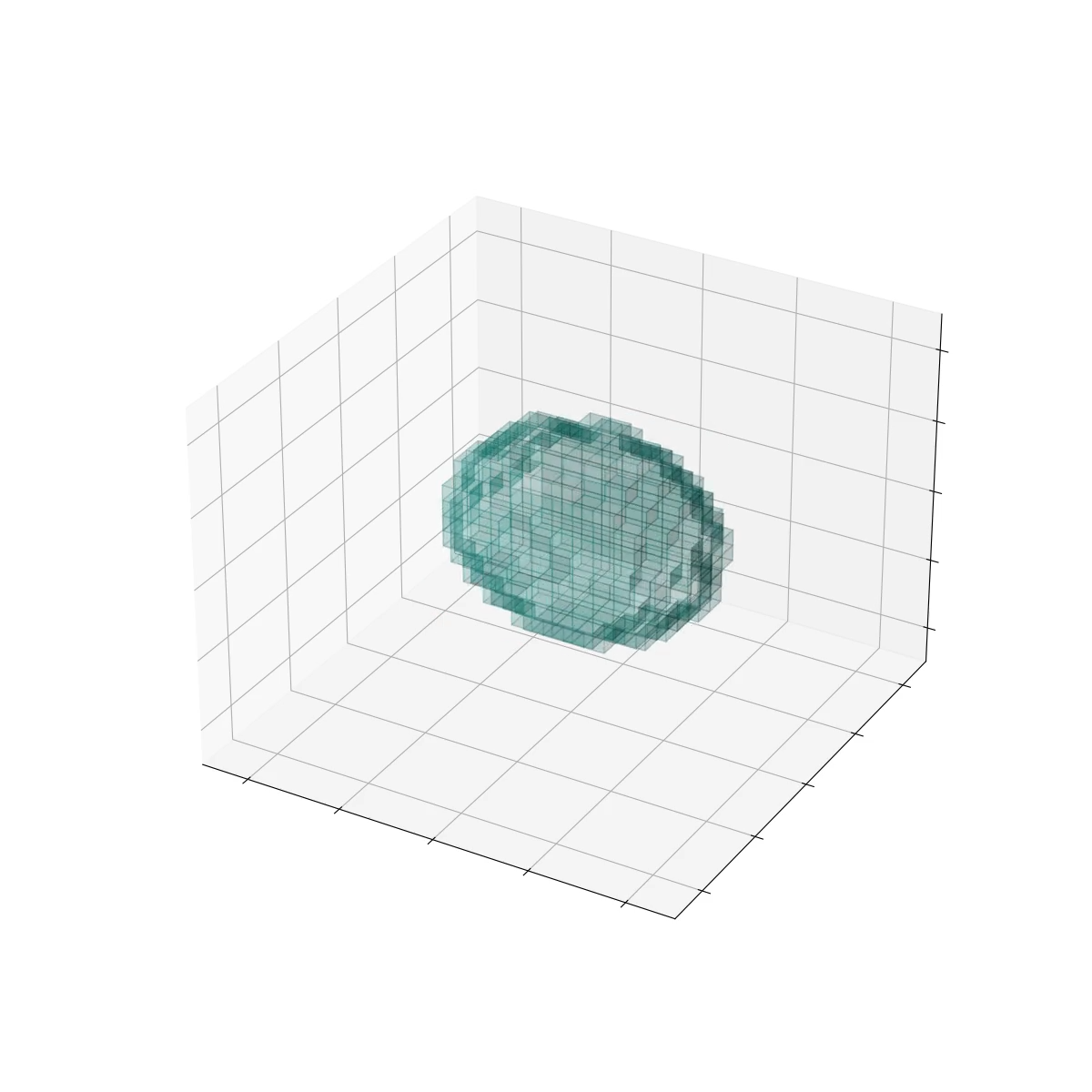}\\
\includegraphics[width=.33\textwidth]{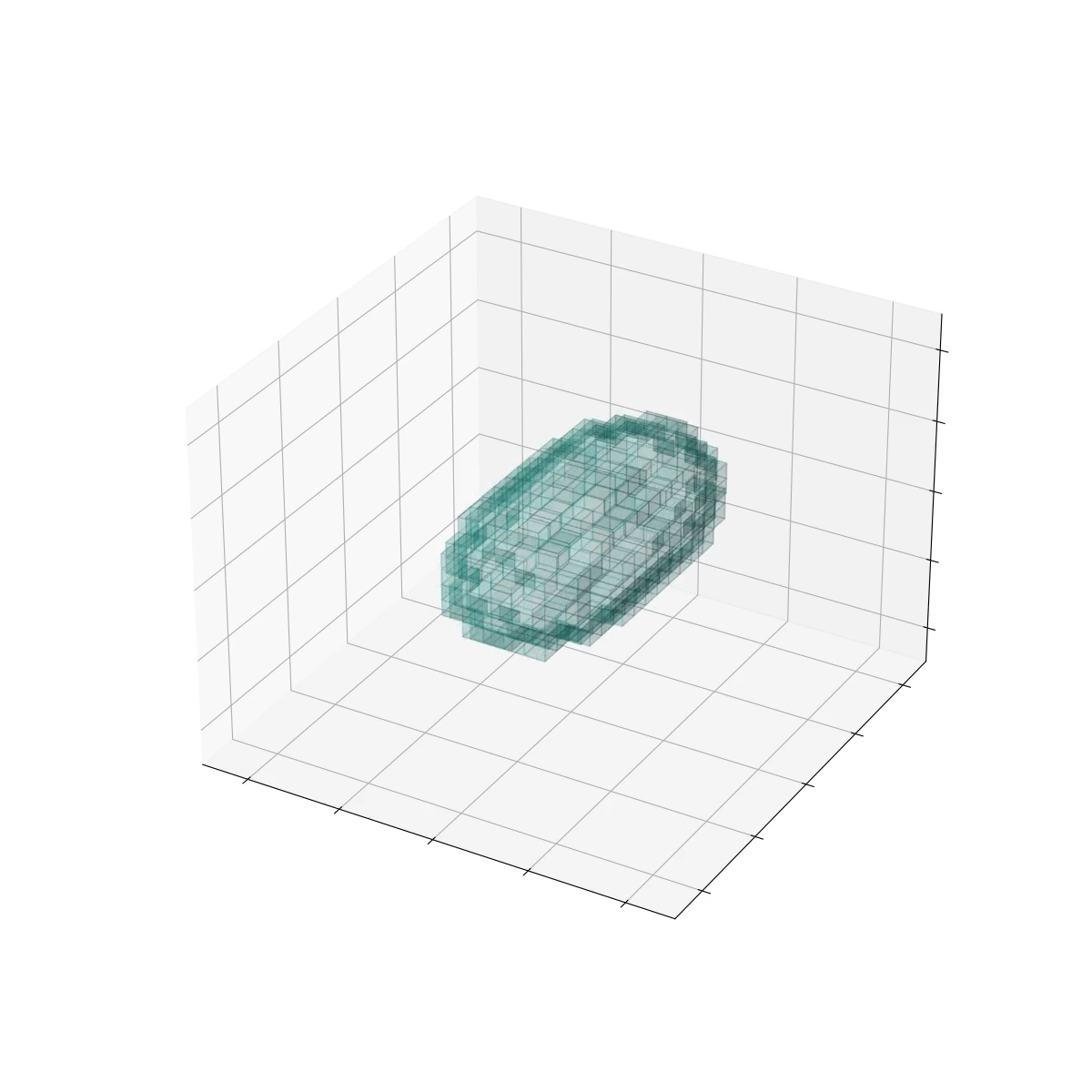}\hfill
\includegraphics[width=.33\textwidth]{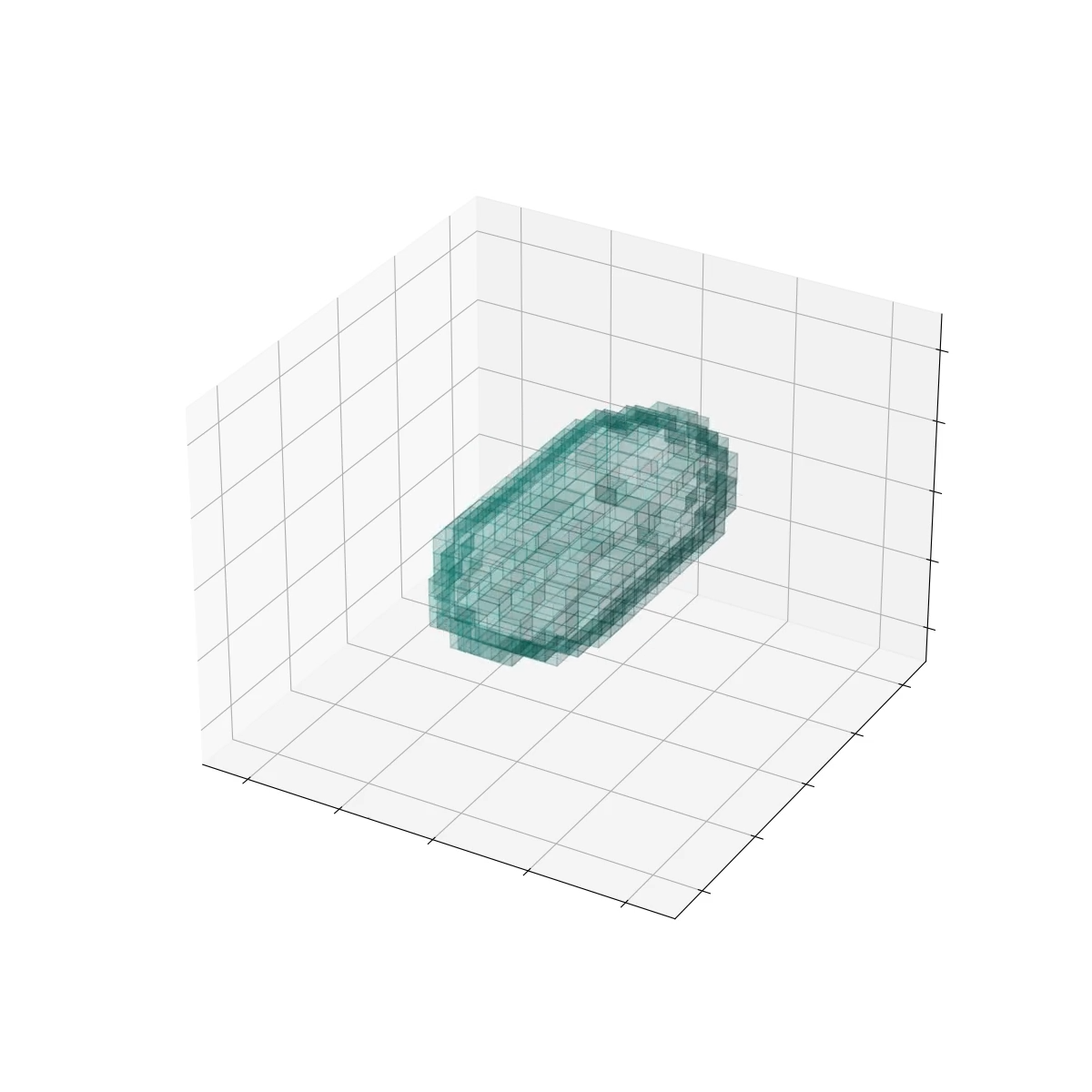}\hfill
\includegraphics[width=.33\textwidth]{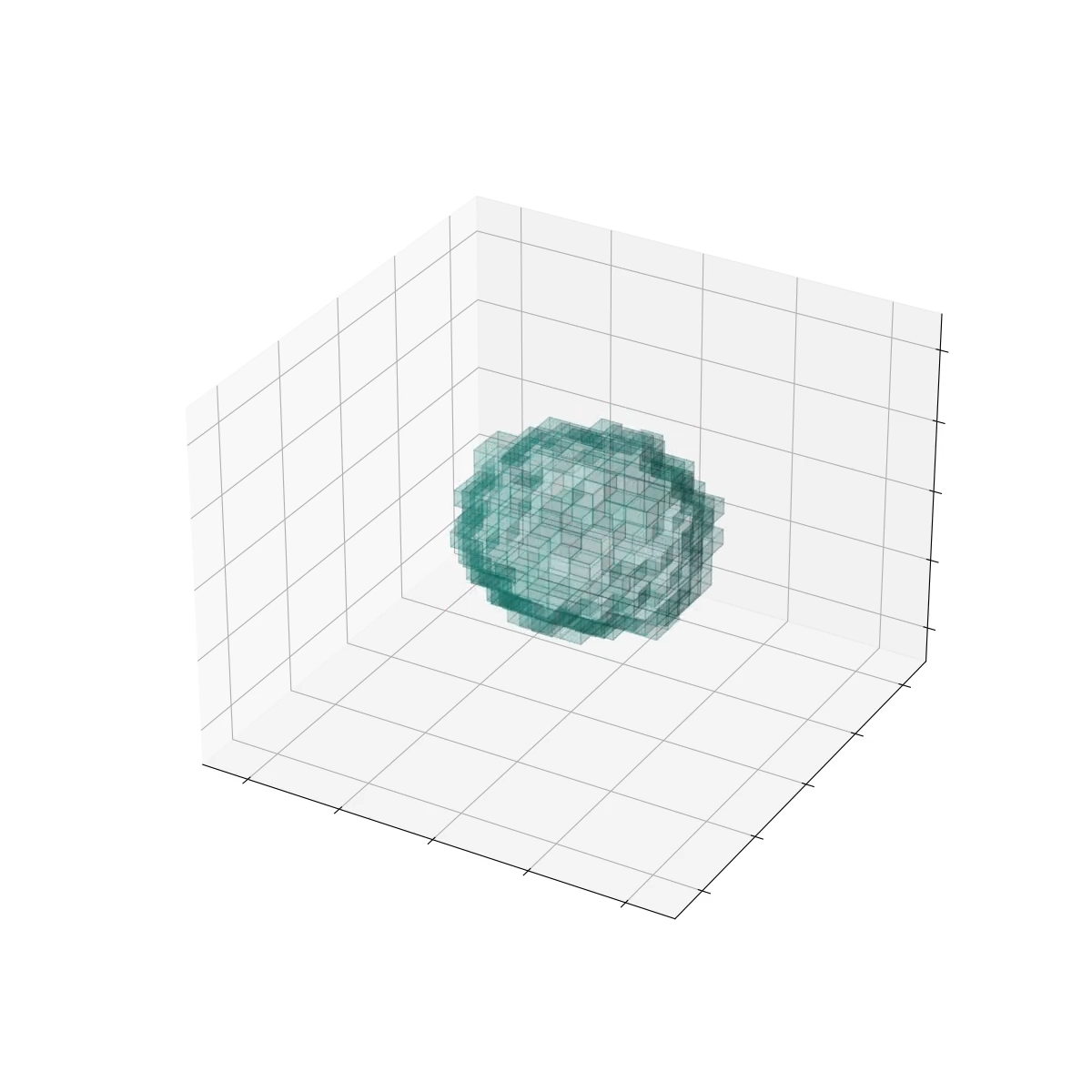}
\caption{Equal charge Q-balls during a $v=0.15c$ collision shown completing one full oscillation cycle in the scalar radial mode, becoming compressed in the axis of the collision before elongating in the same axis before returning to its initial shape. }
\end{figure*}

\begin{figure*}[t]

\centering
\includegraphics[width=.33\textwidth]{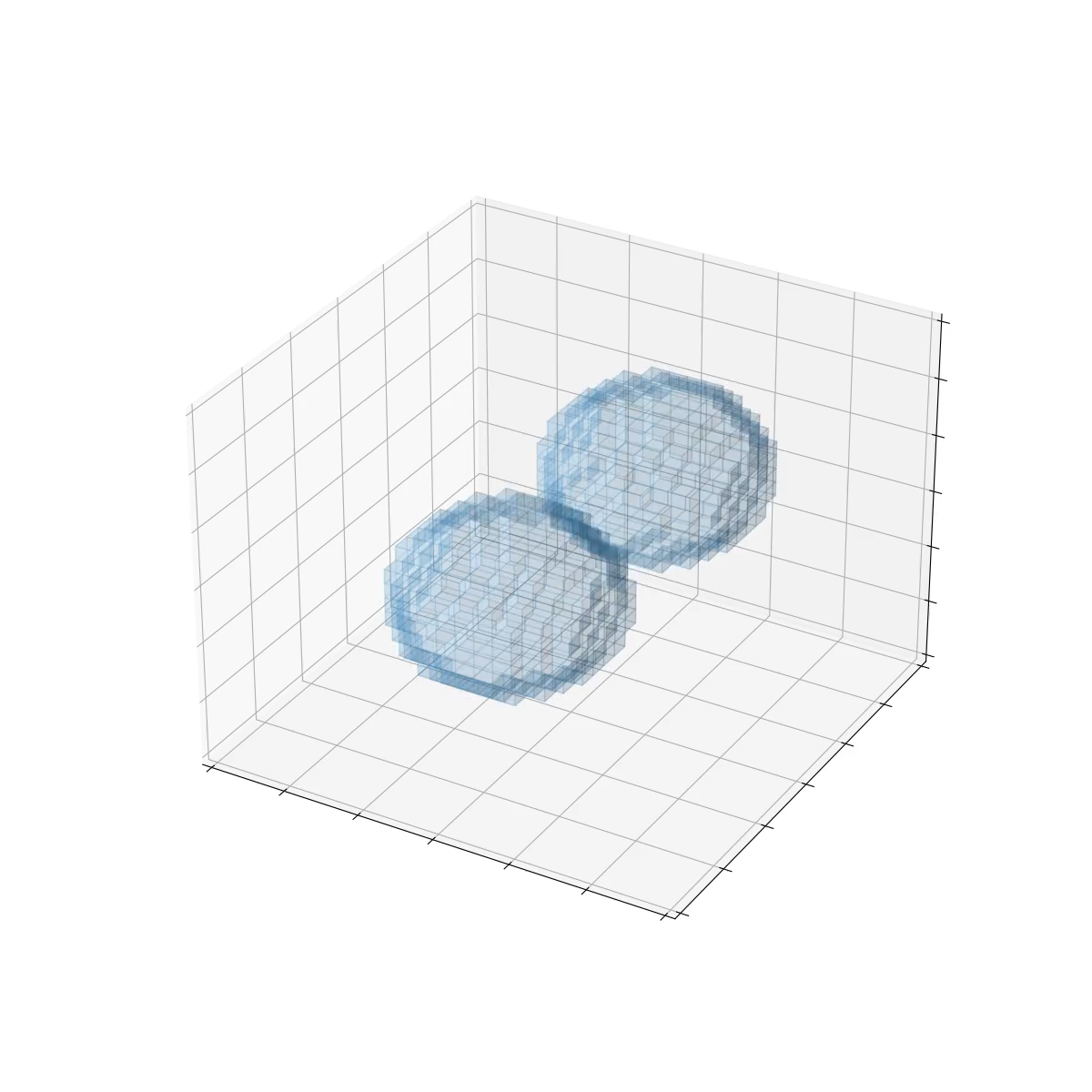}\hfill
\includegraphics[width=.33\textwidth]{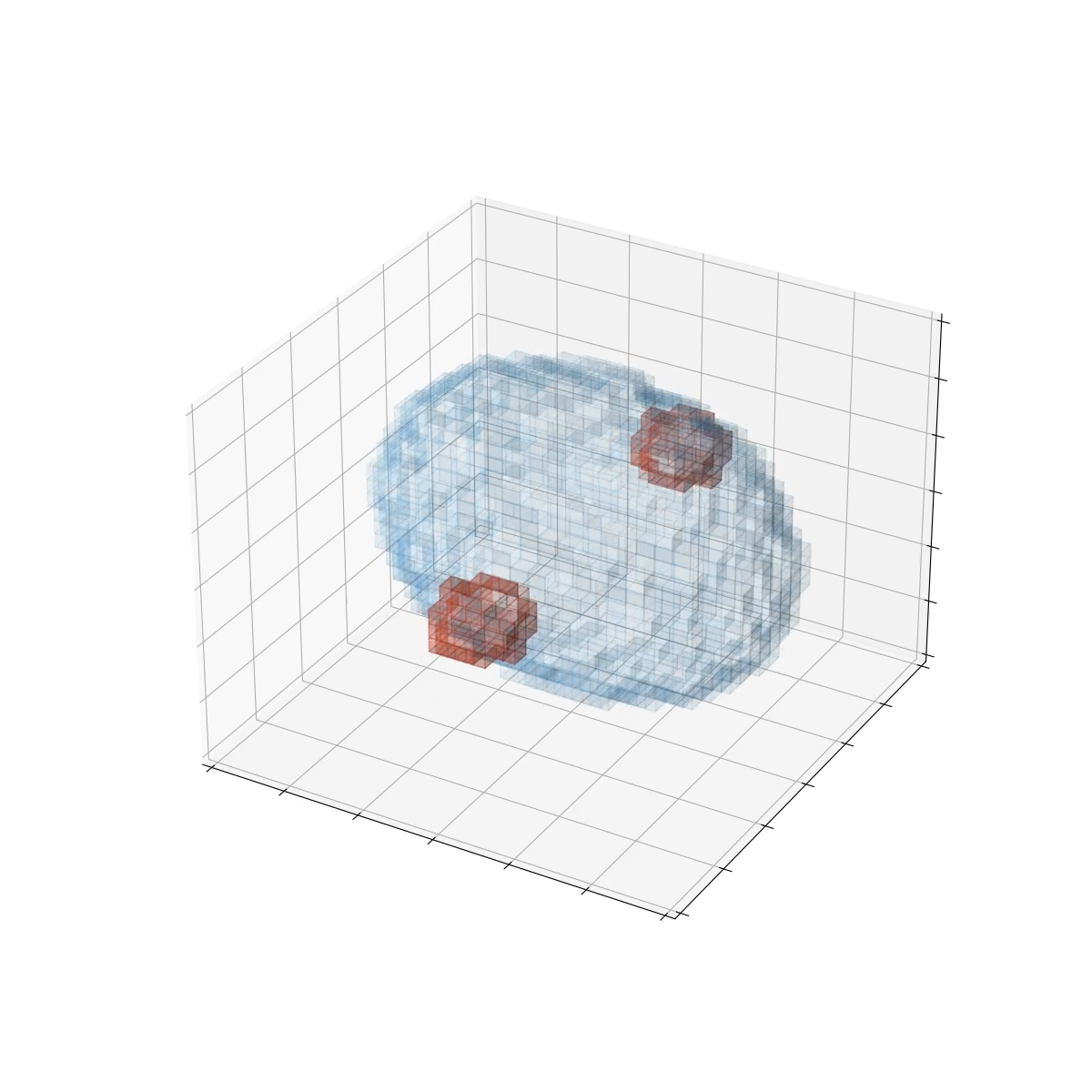}\hfill
\includegraphics[width=.33\textwidth]{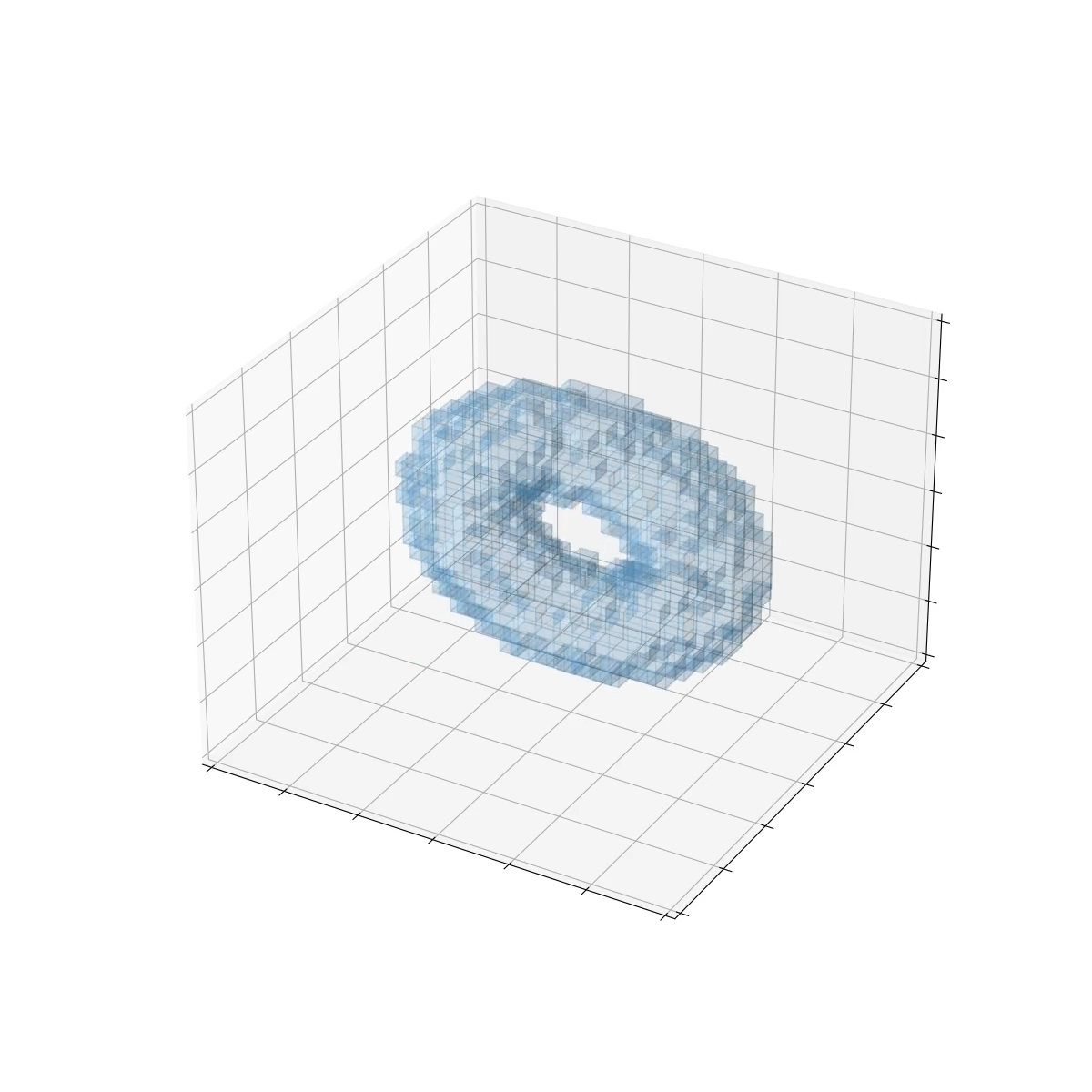}\\
\includegraphics[width=.33\textwidth]{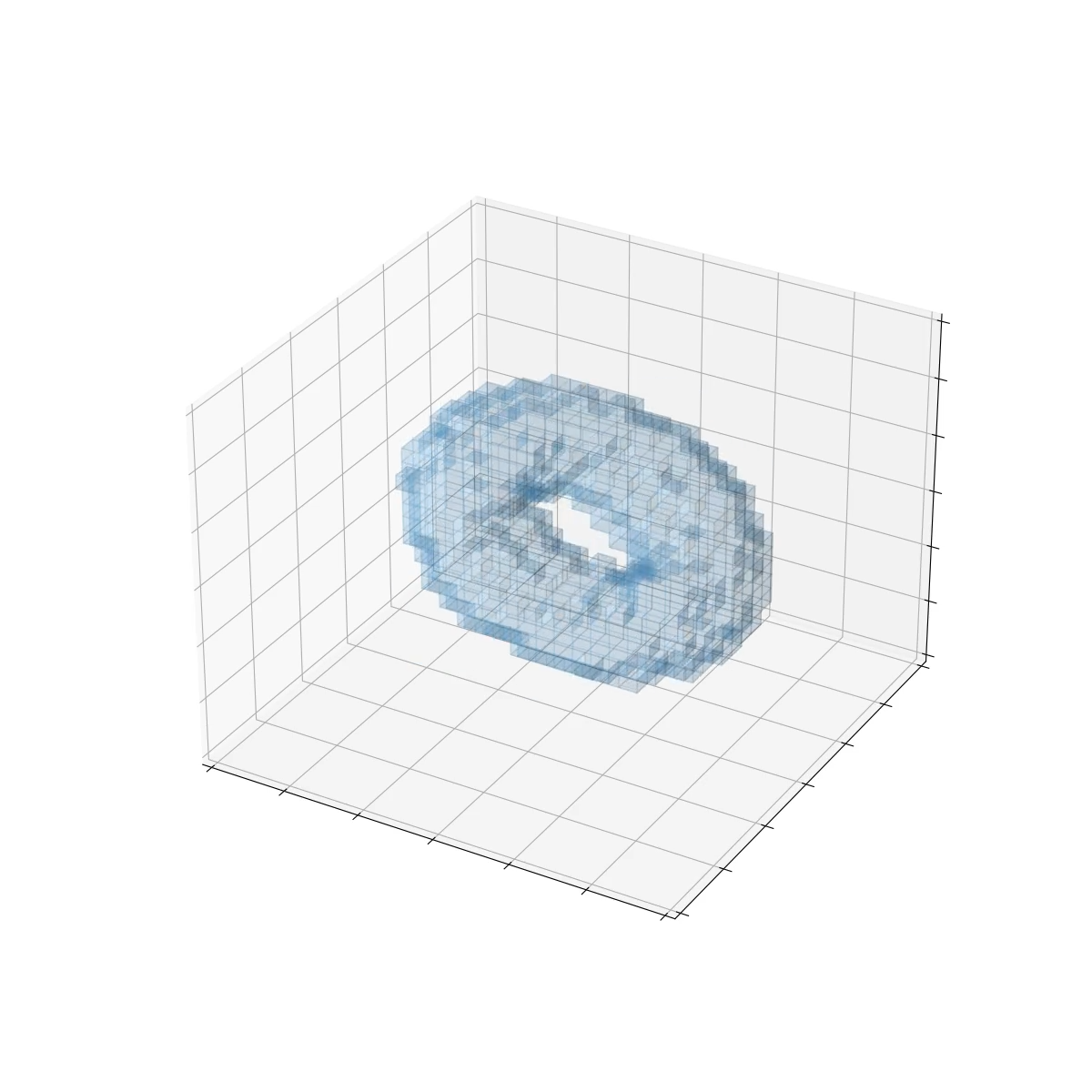}\hfill
\includegraphics[width=.33\textwidth]{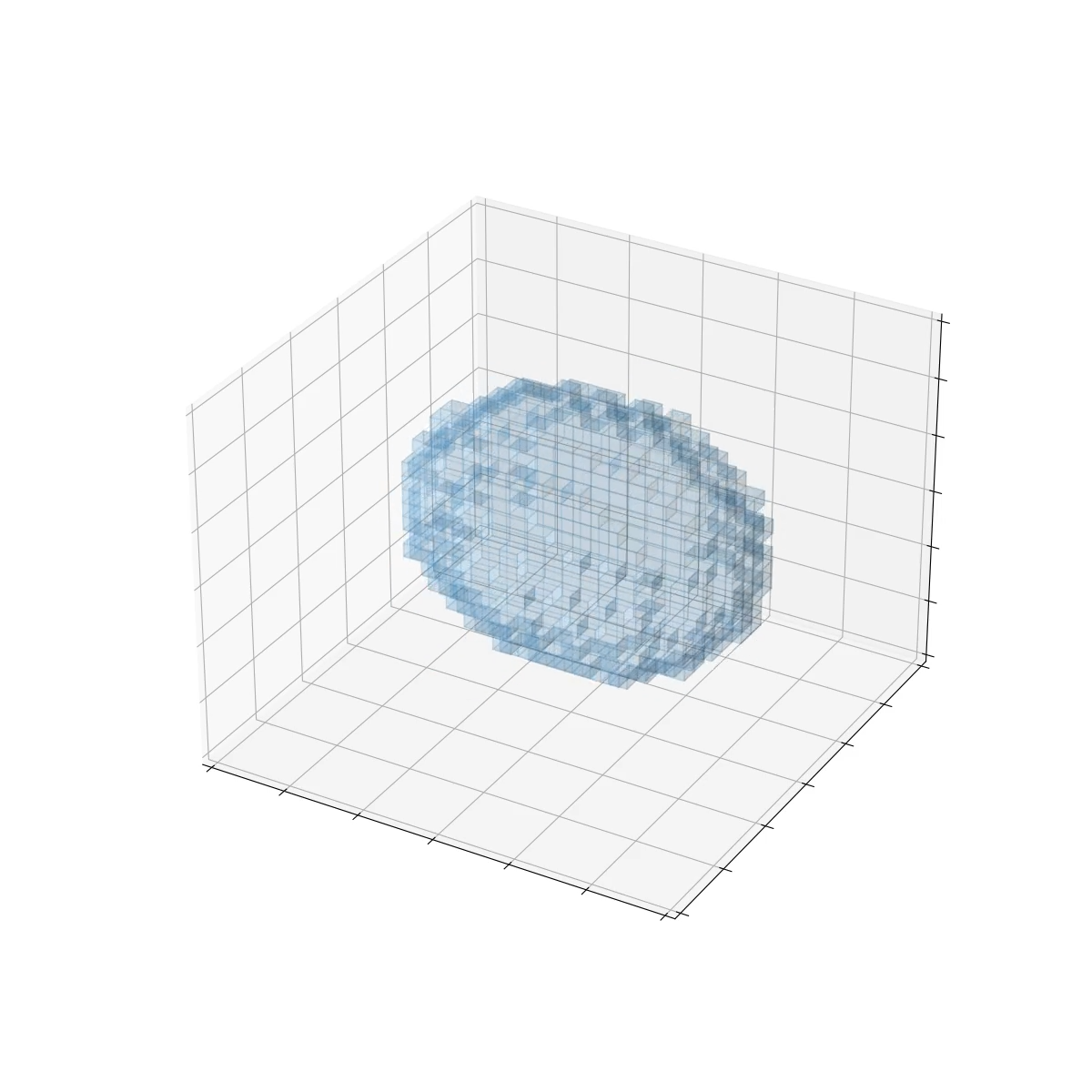}\hfill
\includegraphics[width=.33\textwidth]{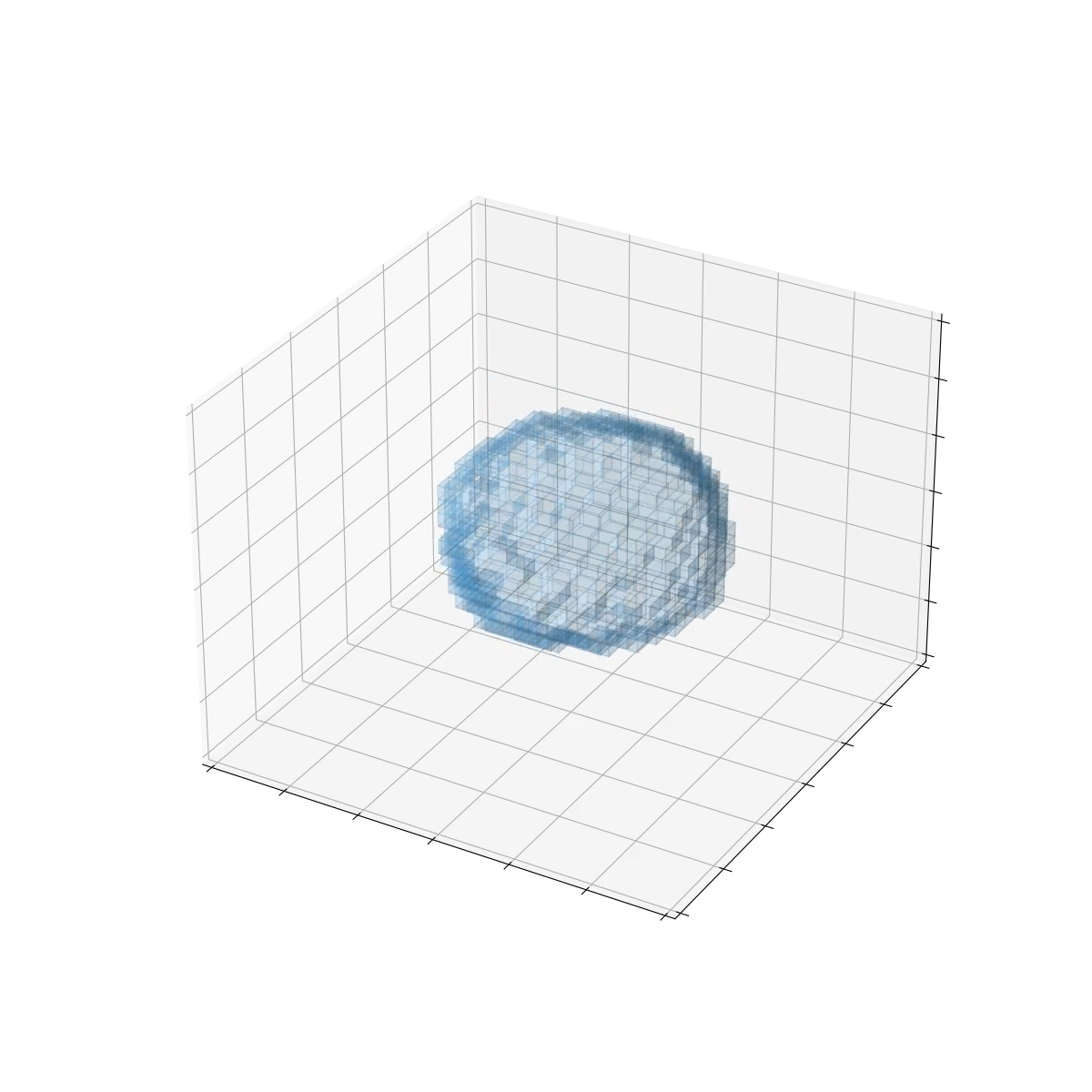}
\caption{Ring collision in which two equal charge Q-balls collide head-on at $v=0.2c$ along the y-axis. After colliding two smaller Q-balls are ejected (in red) while the remaining charge momentarily forms a ring object in the x-z plane. After reaching maximum size, the ring contracts down into an excited state similar to Fig. 4. }
\label{fig:figure5}
\end{figure*}

\begin{figure*}[t]
\centering
\includegraphics[width=.33\textwidth]{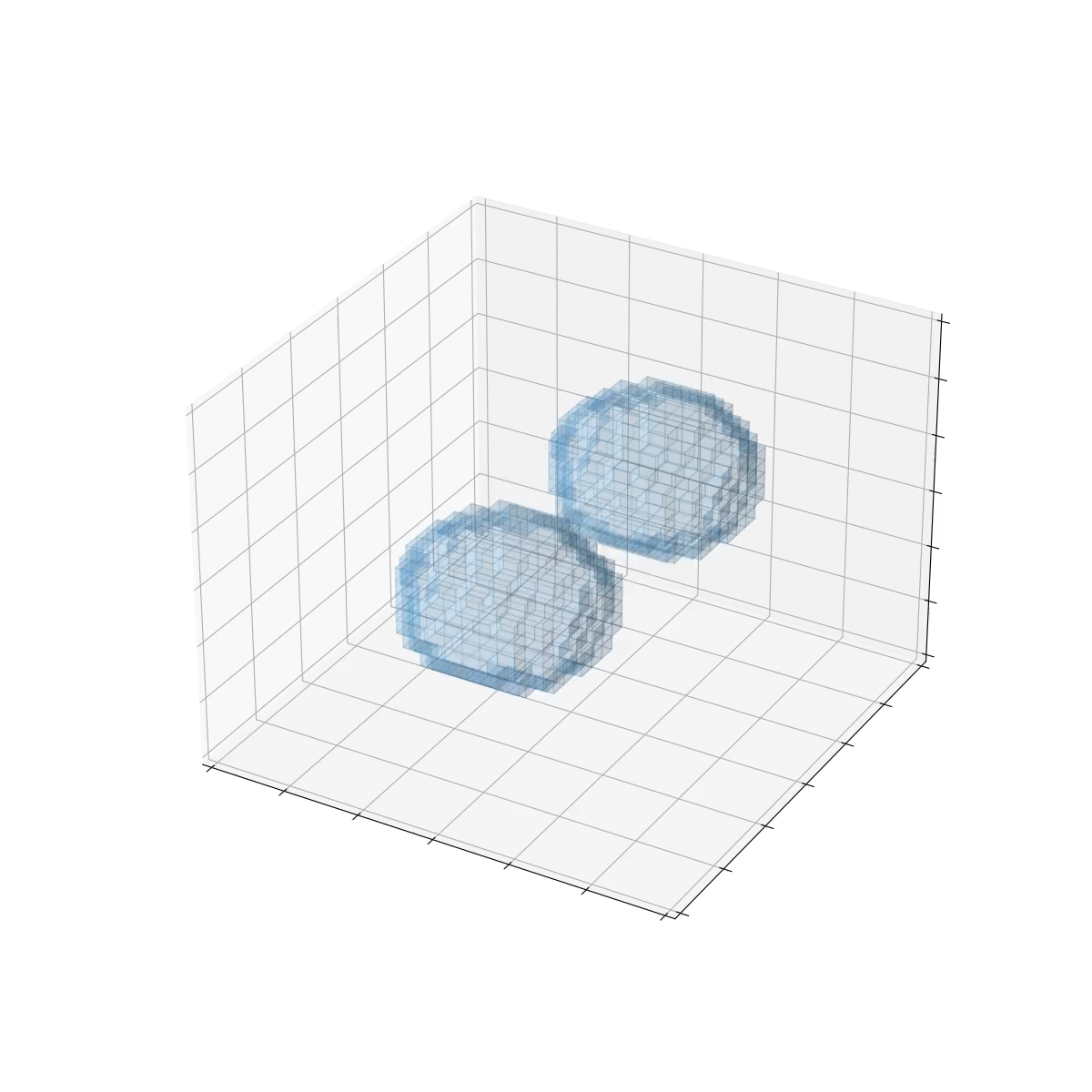}\hfill
\includegraphics[width=.33\textwidth]{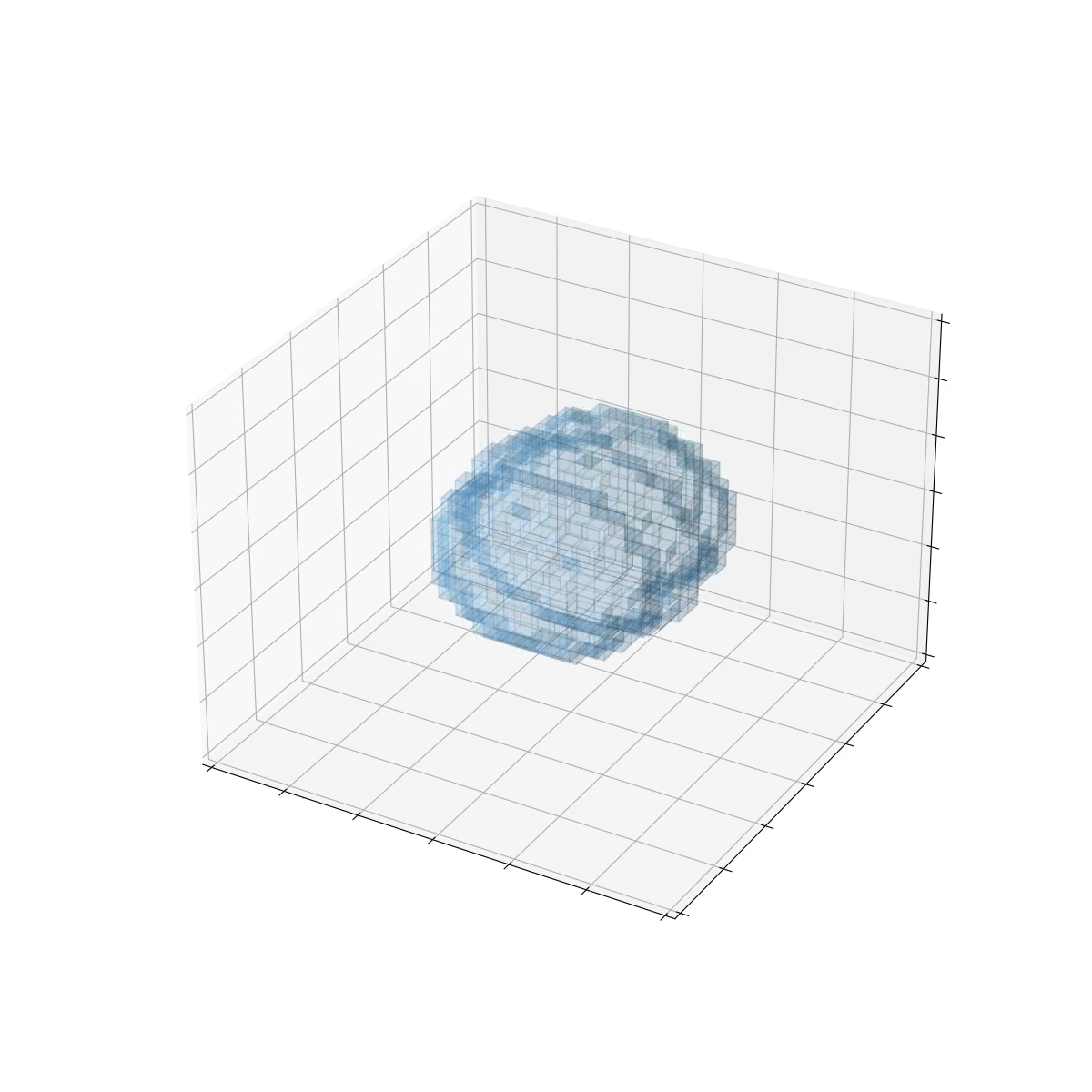}\hfill
\includegraphics[width=.33\textwidth]{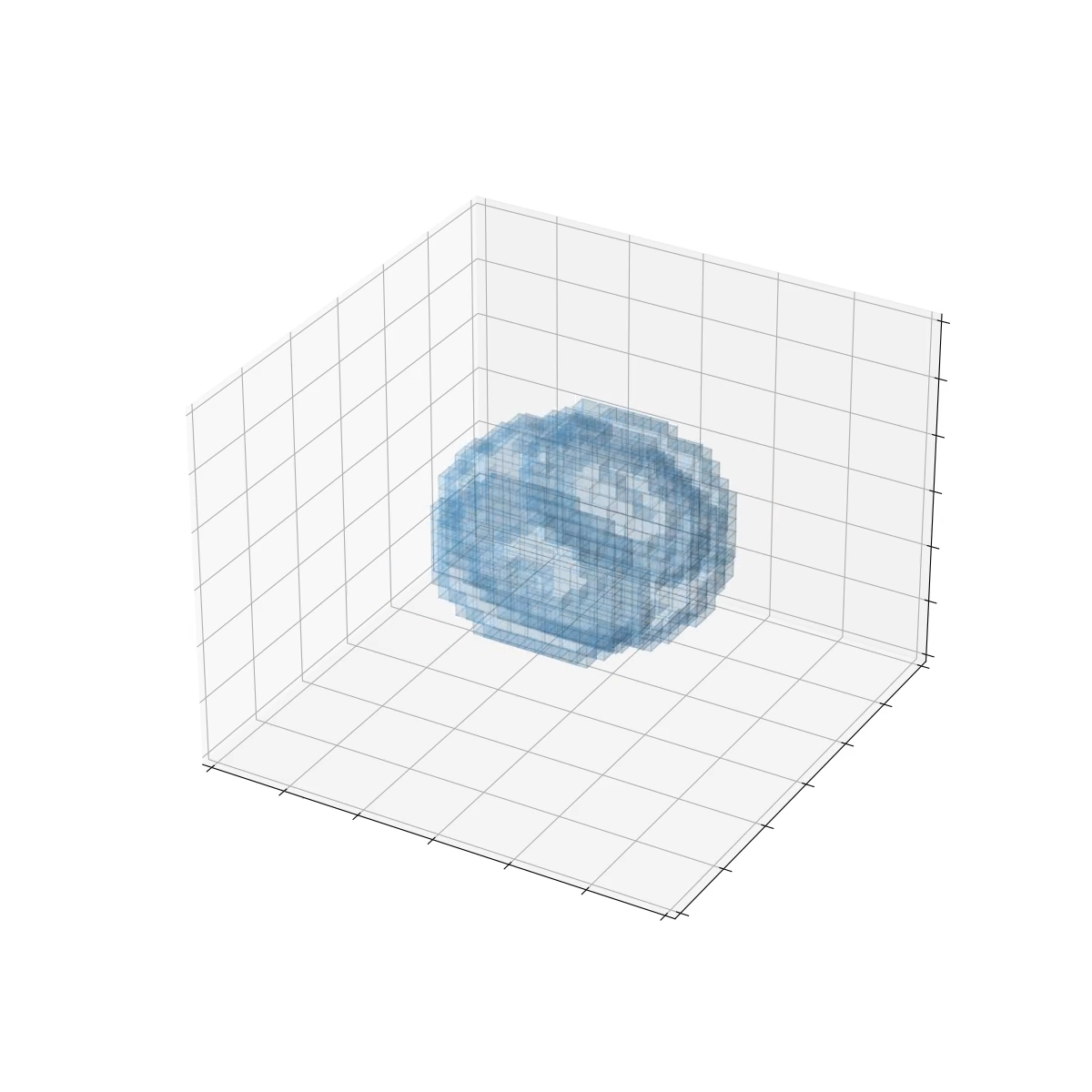}\\
\includegraphics[width=.33\textwidth]{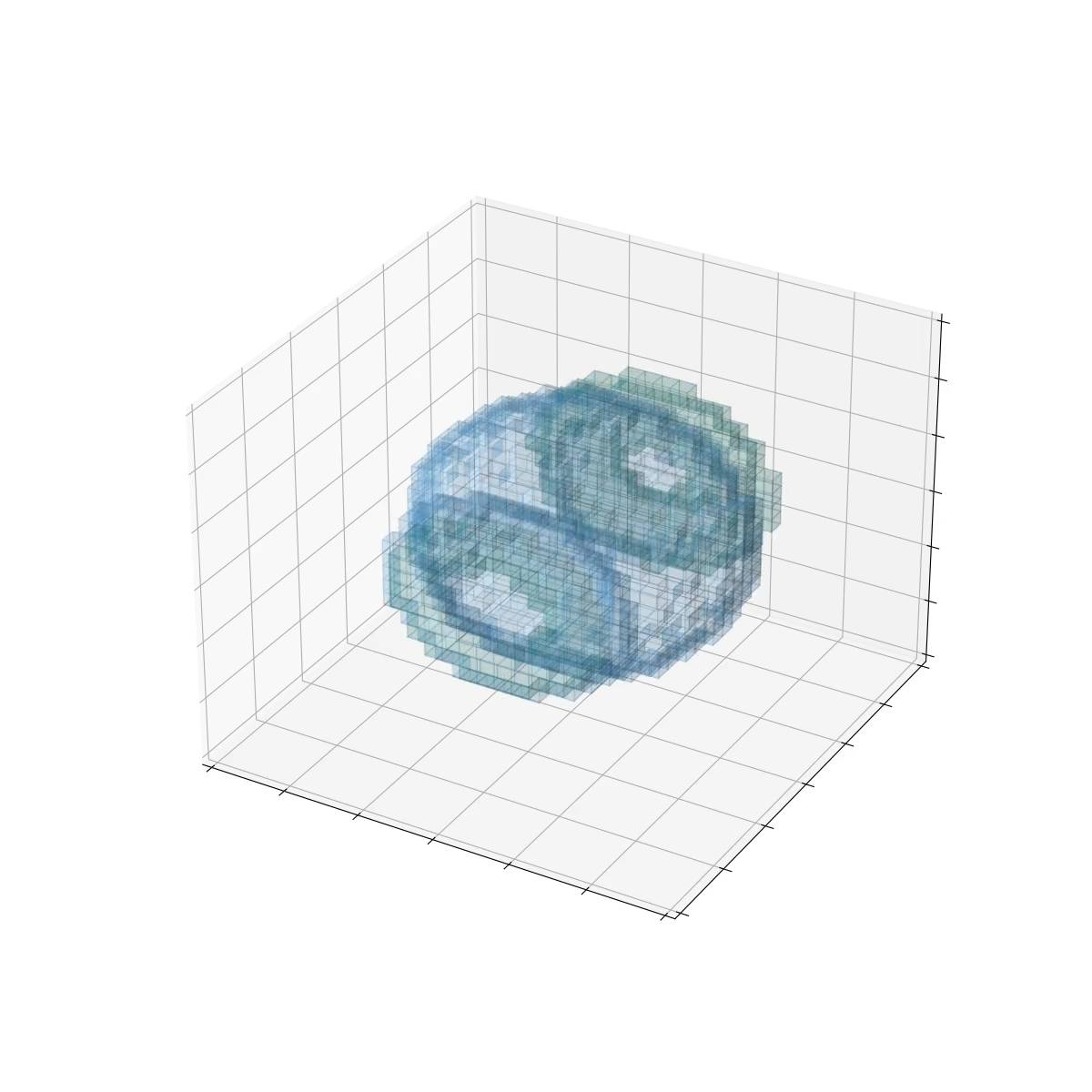}\hfill
\includegraphics[width=.33\textwidth]{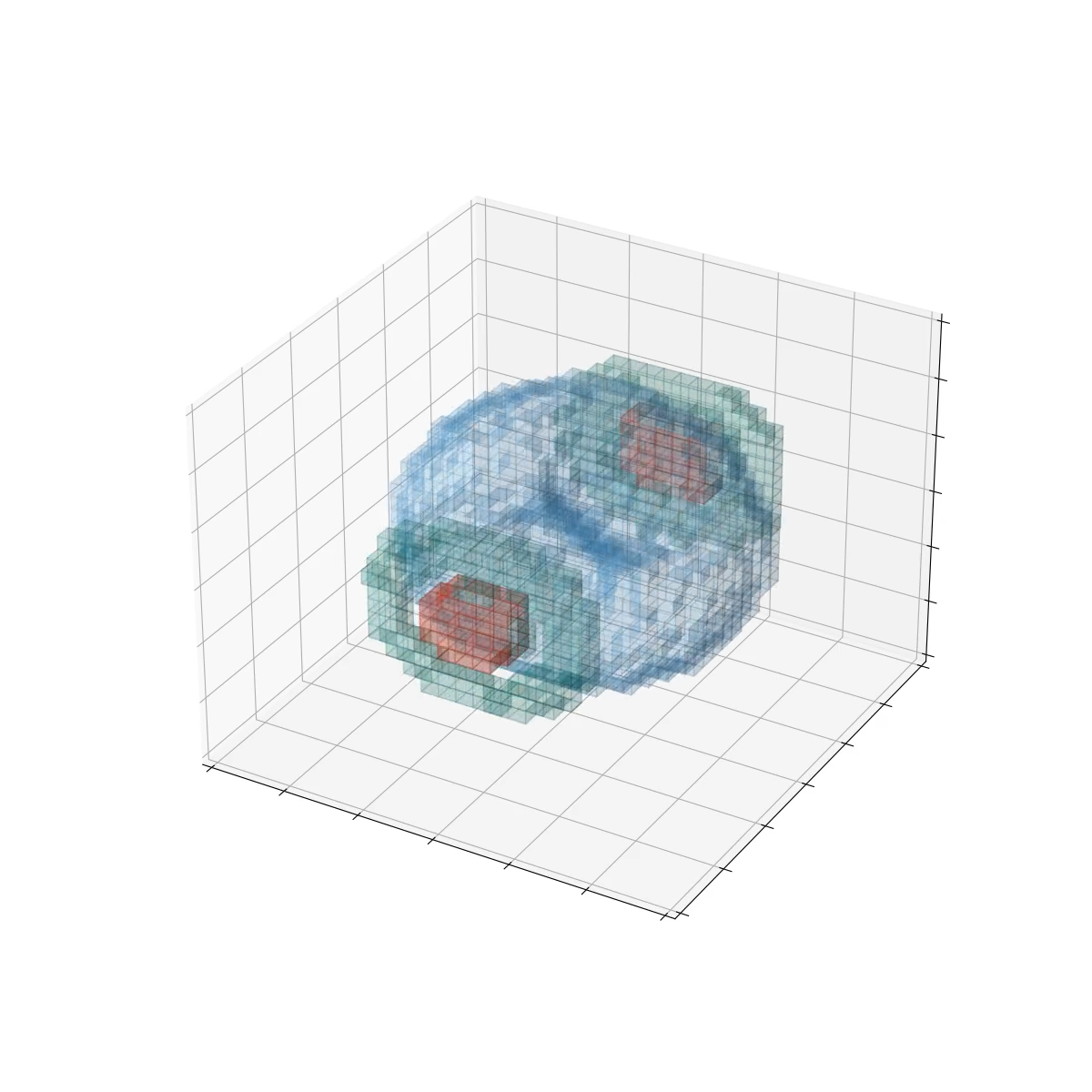}\hfill
\includegraphics[width=.33\textwidth]{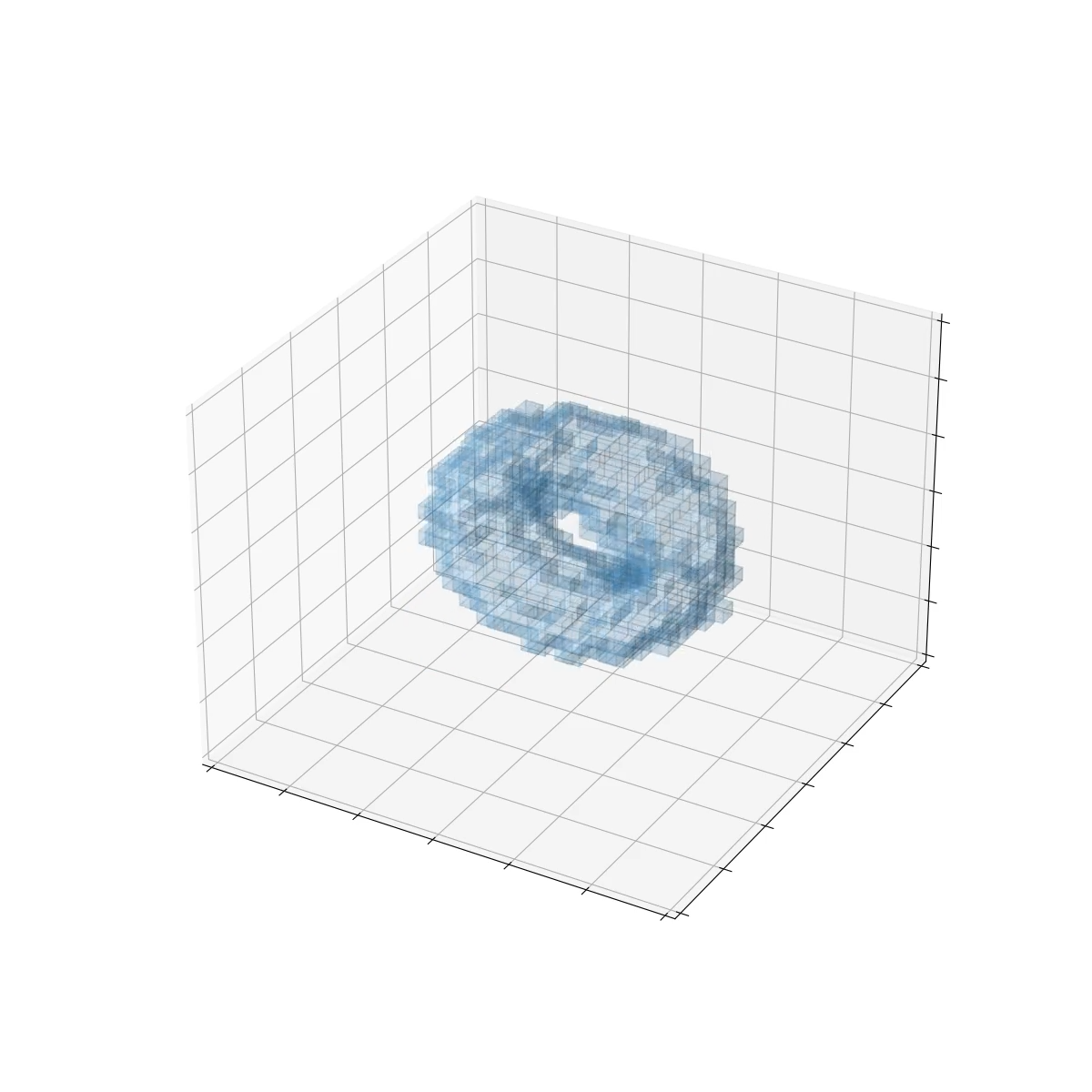}
\caption{Ring collision in which two equal charge Q-balls collide head-on at $v=0.4c$ along the y-axis. After colliding two smaller moving Q-Rings are ejected (in green), followed by two Q-balls (in red) while the remaining central charge forms a ring object in the x-z plane. After reaching maximum size, the ring contracts down into an excited state similar to Fig. 4.}
\label{fig:figure6}
\end{figure*}

\subsection{Initial Boosted State}
For a stationary stable Q-ball we can solve for the field profile via shooting method. This is then applied with our spherically symmetric conditions to a 3D profile on the lattice with radius R set as by the profile. The associated angular frequency is solved for the given profile and value of the field at the origin. In order to find the numerical initial state of a moving Q-ball we boost the solution for a stable stationary Q-ball, similar to the 2D and 3D approaches in Refs. \cite{Gutierrez:2013wiv, Battye_2000}. As analytic formula, the initial variables for a Q-ball boosted by $\gamma=1/\sqrt{1-v_Q^2}$ for a velocity $v_Q$ in units of speed of light ($c=1$) are
\begin{align}
    &\Pi_1=\frac{d\sigma(\gamma x,y,z)}{\gamma dx}cos(\omega\gamma v_Q x)v_Q \gamma-\omega \gamma \phi_2\\
      &
     \Pi_2=\frac{d\sigma(\gamma x,y,z)}{\gamma dx}sin(\omega\gamma v_Q x) v_Q \gamma -\omega \gamma \phi_1\\
      &
    \phi_1=cos(\gamma \omega v_Q x)\sigma(\gamma x,y,z)\\ &
     \phi_2=sin(\gamma \omega v_Q x)\sigma(\gamma x,y,z). &
\end{align}
This is then used to find a numerical lattice solution to the nearest boosted coordinates. For a collision event, two boosted solutions in opposite directions are set to collide in the centre of the comoving box with periodic boundary conditions. In order to set the initial relative phase we can cycle the initial boosted states between the the fields $\phi_1,\phi_2,\Pi_1,\Pi_2$, up to minus signs. That is, the periodic cycling between the two Cartesian component fields inform us as to how, for example, the $\theta_{ph}=\pi$ out-of-phase state relates to the boosted profiles of the $\theta_{ph}=0$ in-phase state solved for initially. Fig. 2 shows an example of initial boosted Q-ball profiles for a head-on collision.

\section{Results}

\subsection{Pair of same charge Q-balls}
Figs.~3 - 6 show three typifying examples of equal charge boosted Q-ball collision simulations in Minkowski-space with potential $V_I$ without gravity plotted. Lattice size for these is $L^3=50^3$, with $M=1$ in the potential setting the units for $A$, and the reduced Plank mass $M_{p}$~\footnote{Note that all dimensional parameters in the paper are in units of M, which can be re-scaled. For the estimation of gravitational energy, however, we take $M=1\, {\rm GeV}$ as an example.}.  These plots show the  energy density of scalar fields over a series of times during the simulation.  These examples are pass-through events at high velocity, merger and ejection of Q-balls at intermediate critical velocities, and finally excited-state creation at low velocity, with radial oscillation frequencies set by the excess energy of the merger. 
For a given potential with potential parameters, Q-ball velocity and Q-ball charge, we can setup any head-on or tangential collision between several Q-Balls.  Modifications can allow for early universe collisions following an AD fragmentation event. Of interest are the final state distribution of charge, the excited states, the limit of Boson-stars and gravitational binding energy for final states, the emission of gravitational waves and relevent energy loss.
We focus on highly-boosted initial states that are not gravitationally self-bound in their stationary frame.



\subsection{Q-Ball and Anti-Q-Ball}
We perform two simulations of head-on Q-ball and anti-Q-ball collisions, one each with a relative phase difference of $0,\pi$ respectively. Here the energy dissipated in the scalar field is accompanied by a spherically symmetric outgoing gravitational wave (GW) signal. Fig.~7 shows a Q-ball and anti-Q-ball with a relative phase difference of $\pi$ and the associated scalar radiation can be seen in the scalar energy component modulations. 
Two smaller Q-balls of opposite sign are ejected in this case with no central object.
Fig.~8 shows the case without phase difference, $\theta_{ph}=0$ and one sees similar signals, though, with more charge violation in the simulation. This case also resolves to a Q-ball and anti-Q-ball final state. Due to numerical errors, these collisions in Figs. 7 and 8 feature tiny transient violations of energy conservation produced by very large field amplitudes from the energetic annihilation of ground sates. Figure 9 shows the charge density for a tangential collision at $v=0.4c$ and the angular momentum of a combined object in the immediate aftermath of colliding. 

\subsection{Critical Velocity}
For relative velocities in a critical range, the head-on collisions of Q-balls exhibit interesting features as shown in Figs. 5 and 6. In Fig.~5 the central object ejects two smaller Q-balls while the larger central mass forms a Q-ring before collapsing to an excited state. On the other hand, if the relative velocity is above the critical range, the Q-balls just pass through each other, seen in Figs. 3 and 10,  while for smaller relative velocities they coalesce, as in Figs. 4 and 11.

In \cite{Axenides:2001jj} the formation of Q-rings is noted as in principle possible in any Hamiltonian that admits Q-ball solutions. The mixture of topological and non-topological charge at work, however, makes it more difficult to find these solutions analytically than Q-ball profiles.  Fig.~12 one of the metric perturbations is plotted with a Q-ring collision simulation in order to observe the impact of the critical velocity events on the asymmetry of the metric perturbations during the formation of the ring. Fig.~6 shows a triple ring formation with two smaller ejected Q-balls, an outcome also seen in the 3D simulations in Ref. \cite{Battye_2000}. The same event is analysed in Fig.~13. The period of oscillations in one of the GW signal components for Figs. 12 and 13, which are, though, absent in the scalar energy figure, are taken to be spatial variations associated with excitations above ground state Q-balls and rings, with energy determined by the difference between ground and excited states.



\subsection{Newman-Penrose Scalar}
One of the relevant variables for the GW signal from the metric data is the Newman-Penrose (NP) scalar, $\Psi_4$ \cite{Newman:1961qr} extracted from spin-weighted spherical harmonics, ${}_{s}{Y_{lm}}$, and full NP scalars over the 2-sphere at each radius around the collision site we have
\begin{equation}
\Psi_4(t,r) = \int_{s_2} {\Psi_4^{lm}}(t,r)\, \, {}_s{Y_{lm}}\,  d \Omega,
\end{equation}
where the flux of radiated energy can be written as
\begin{equation}
\frac{dE}{dt} = \lim_{r\to\infty} \frac{r^2}{16 \pi} \sum_{l,m} \left| \int_{-\infty}^t \Psi^{l,m}(t') dt' \right|^2,
\end{equation}
and the NP scalar is found by contracting the Weyl tensor,
\begin{align}
C_{\mu \nu \rho \sigma} &= R_{\mu \nu \rho \sigma} \nonumber \\&+\frac{1}{n-2}(R_{mg} g_{kl}  - R_{il} g_{km}+R_{kl} g_{im}-R_{km} g_{il}) \nonumber  \\&+\frac{1}{(n-1)(n-2)}R (g_{il}g_{km}-g_{im}g_{kl}),
\end{align}
with a suitably defined null tetrad,
\begin{equation}
\Psi_4=C_{\alpha \beta \gamma \delta} k^{\alpha}\bar{m}^\beta k^\gamma \bar{m}^\delta.
\end{equation}
The NP scalar accounts for the gravitational wave energy carried away to infinity.
For very distant observers at at large distance $r_o$ the relevant signal using the stress energy tensor is 
\begin{equation}
\bar{h}_{ab}= \frac{2 G}{r_o}\ddot{I}_{ab} (t_r)
\end{equation}
where the quadrupole moment is defined by the time-time component of energy-momentum tensor, $T^{00}$,
\begin{equation}
I_{ab}= \int d^3 x^a x^b T^{00}(x)\,,
\end{equation}
and can be found from the scalar field energy density data of the simulation.
For the different regimes of size and together with choices of early or late universe one can choose different approaches for the gravitational field e.g. as the relative size or energy density by boosting approaches that of Boson-stars and the over-densities become comparable to those in the cases of Boson-stars and their collisions, which may result in black hole formation. 

For specific models of Q-balls in the early or late eras, the gravitational wave detection prospects will depend on the frequency and density of collision events, as well as the distribution of collision energies. This may also be impacted by collisions with Q-balls that have accumulated in stellar objects~\cite{Kusenko:1997it}.
For the simulations presented in this work, we only consider weak enough field regimes that Minkowski space is the background metric in the region surrounding the collision site and present the gravitational wave signal using Eq.~28.

In Figs. 7, 8, 10 - 13 we plot for each of our six head-on collision cases the outgoing GW signal and the modulation of components of scalar energy over the runtime of the simulation with the conservation of Noether charge Q. In each case the simulation ends once accumulated errors in charge or energy conservation surpass $10\%$ which can also cause the growth of non-physical modes such as in the growing gravitational mode towards the end of the simulation in Fig. 11. Figure 14 shows the variation in one of the metic perturbation conjugate momenta for the same collision as Fig. 5.

We find the gravitational energy from the Q-ball collision depends on the relative velocity and $\omega$ of Q-balls. We expect an increase in GW amplitude for higher velocity and bigger charge, when the collision remains critical at such velocities. Larger charge will also increase the peak frequency. The range of peak frequencies for Q-balls we are considering is found to be around $10\,  \text{Hz}$. The full range of possible signals may be explored with variation of model parameters, however the detailed analysis is left to a future work.

\begin{figure}[H]
\centering
\includegraphics[width=.90\columnwidth]{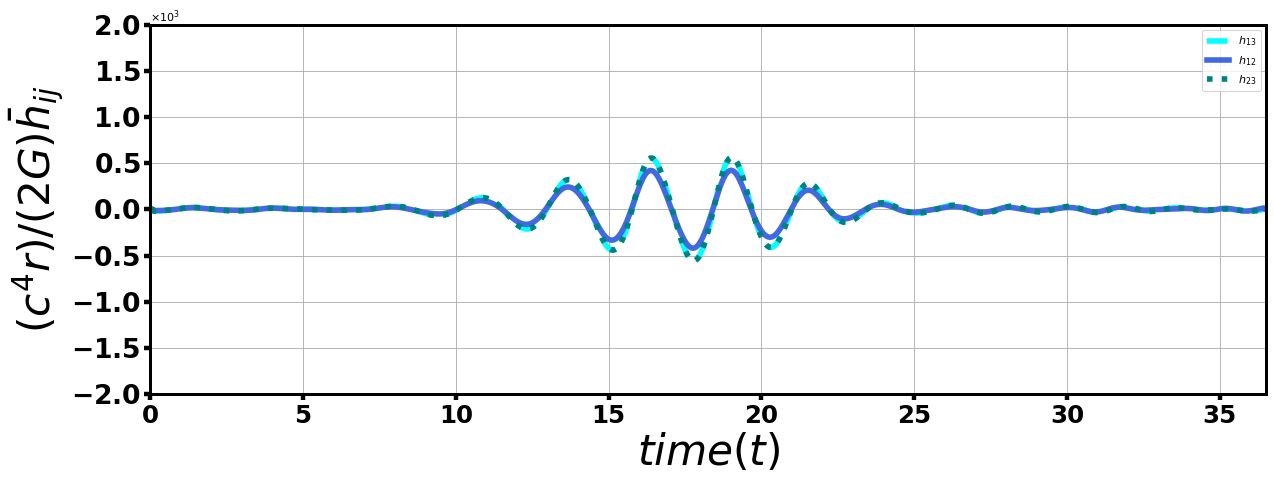}\\
\includegraphics[width=.90\columnwidth]{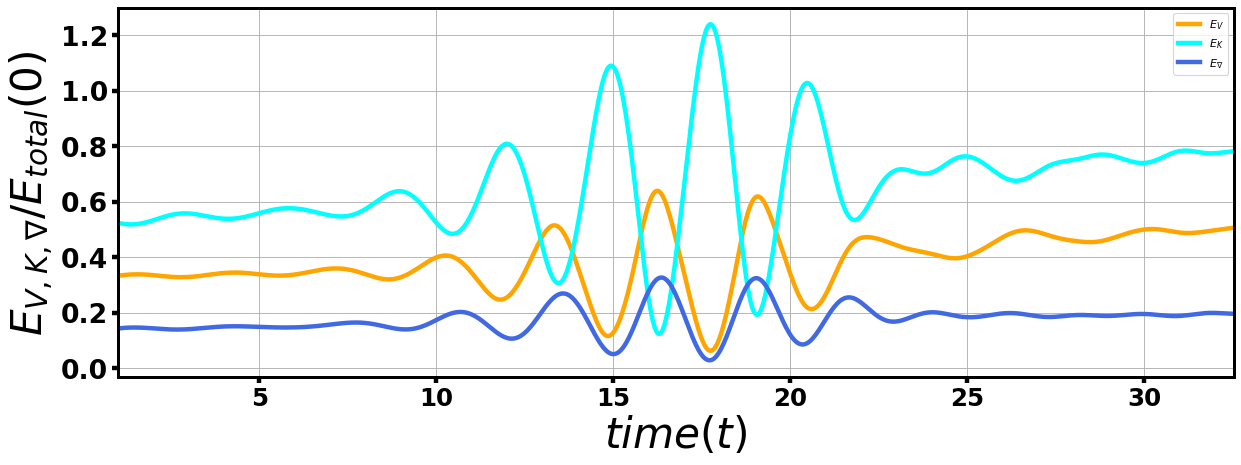}\\
\includegraphics[width=.90\columnwidth]{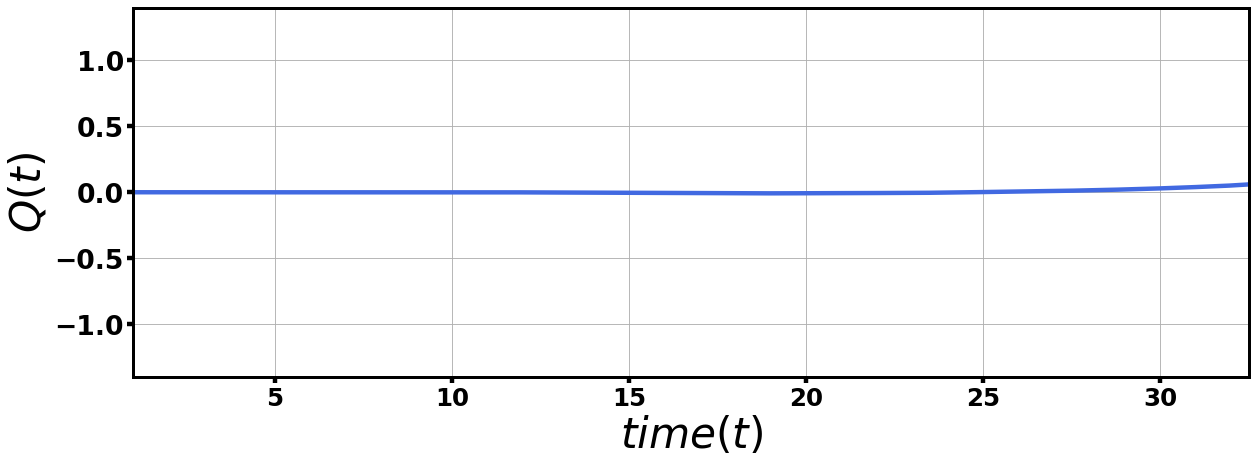}
\caption{Q-ball/anti-Q-ball collision ($\omega^2=0.847^2$, $v=0.3c$) at $t\simeq7$. with relative  phase $\theta_{ph}=\pi$. The collision dissipates scalar field radiation and gravitational energy. Two smaller Q/anti-Q balls are ejected in the aftermath.}
\label{fig:figure7}
\end{figure}

\subsection{Numerical Accuracy}
The length of simulations with high accuracy may be improved with a larger lattice and small $dt$. This may be necessary for examining Q-balls in very long lived states. Figure 15 shows the relative improvement of $\approx 10 \%$ from increasing to larger lattice size, at the cost of longer runtime.

\section{Conclusion}
We have constructed a general toolkit for examining the gravitational waves of Q-ball collisions up to the break down of the weak gravity regime. Examining a number of cases for collisions between Q-balls we have found the expected broad case of outcomes for different relative velocities and studied the gravitational wave signal in each case. 
\begin{figure}[H]
\centering
\includegraphics[width=.900\columnwidth]{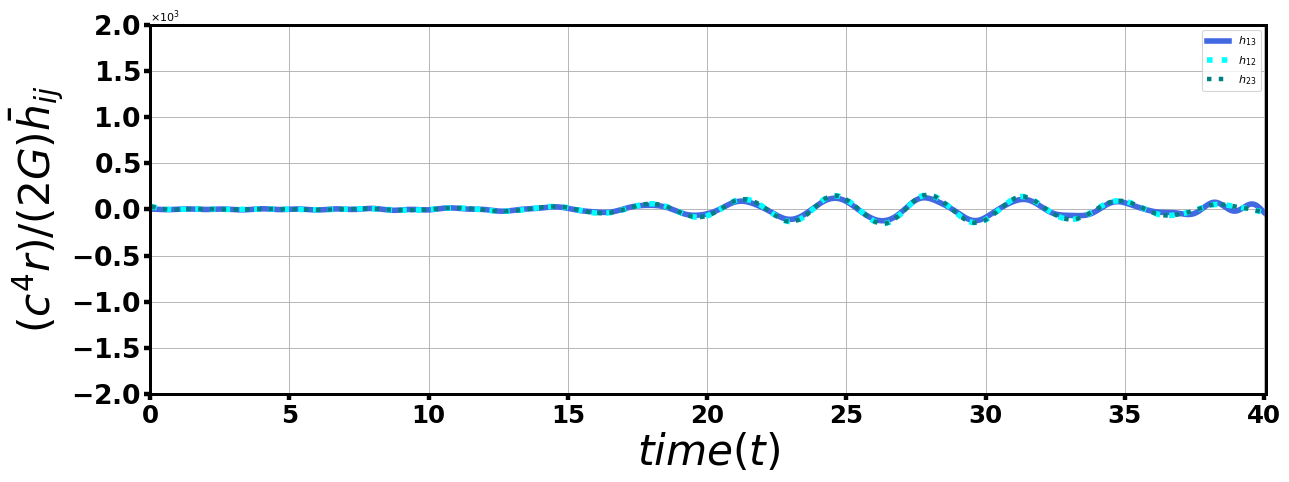}\\
\includegraphics[width=.9000\columnwidth]{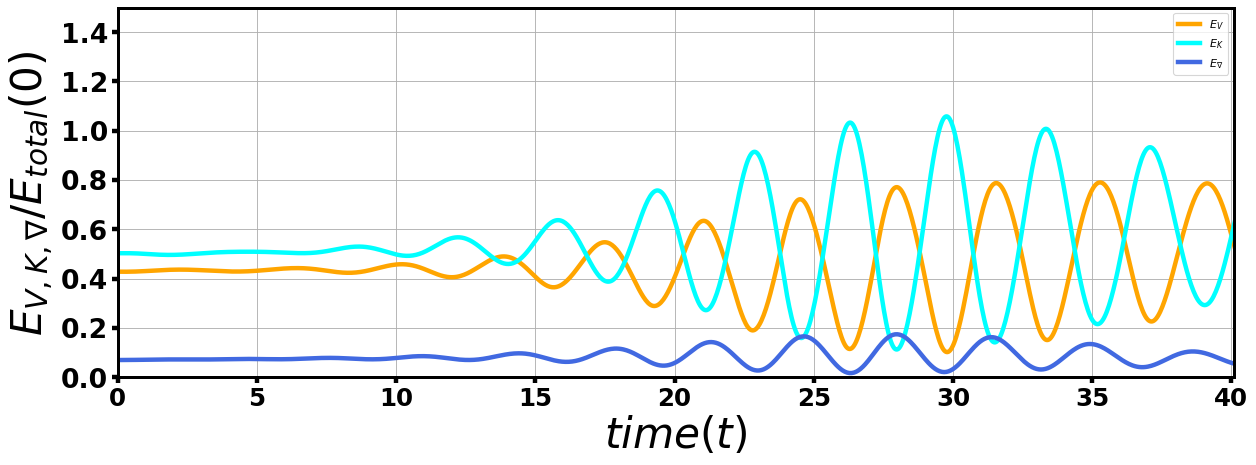}\\
\includegraphics[width=.9000\columnwidth]{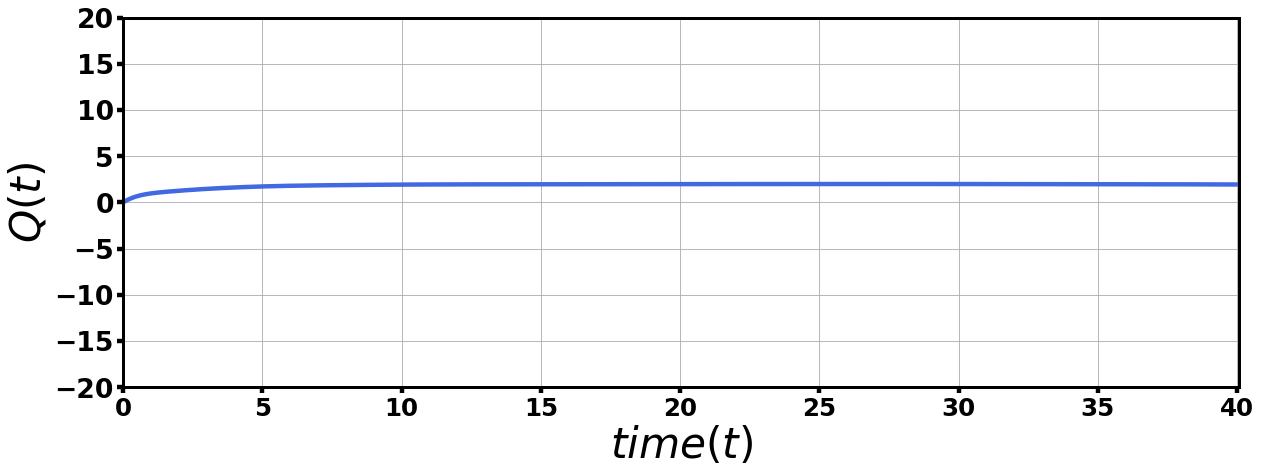}
\caption{Q-ball/anti-Q-ball collision ($\omega^2=0.847^2$, $v=0.3c$) at $t\simeq7$ with relative  phase $\theta_{ph}=0$.}
\label{fig:figure8}
\end{figure}

Annihilating collision events display the largest and most symmetric GW signals, while low velocity excited state collisions and pass through events display weaker signals. Radial mode excited state cycles and their relevance to ongoing gravitational cooling is however important in the long term as opposed to the short time interval of the collision event. For critical velocity cases, the broader range of possible ejected objects and their excitations may yield unique signals that could be identifiable with future GW detectors.

\begin{figure*}[th]
\centering
\includegraphics[width=.65\textwidth]{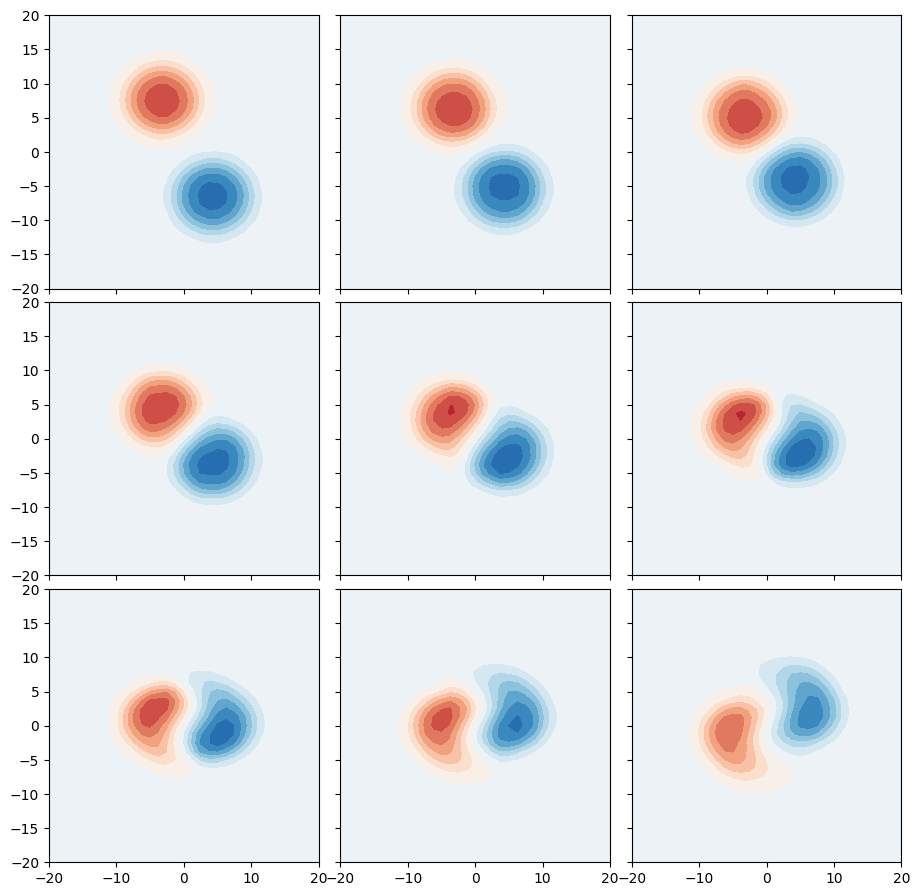}\hfill
\caption{Charge density contour plots showing a tangential collision between a Q-ball (in blue) and anti Q-ball (in red) with equal and opposite charge, $\omega^2=0.806^2$, $v=0.4c.$ After colliding the combined object rotates around the center of mass with a region of zero charge in between, where the two Q-balls are in contact.}
\label{fig:figure9}
\end{figure*}

Beyond the specific cases studied, the variation of the critical velocity with Q-ball size, potential choice and their parameters may be studied with the public release of the code. 
Future work may extend the possibilities to higher density lattices and include cases of the complex scalar coupled to additional light scalar fields such as axions, the coupling to fermions or additional vector fields. 
Early universe collision dynamics shortly following a Affleck-Dine fragmentation are also of interest with suitable adjustments to the metric evolution during the early universe era.

Future work on model specific applications may allow one to examine the detectable signals, per model of BSM physics, for Q-balls in late of early eras.

\acknowledgments
This work was supported by a 2-Year Research Grant of Pusan National University.

\begin{figure}[H]
\centering
\includegraphics[width=.90\columnwidth]{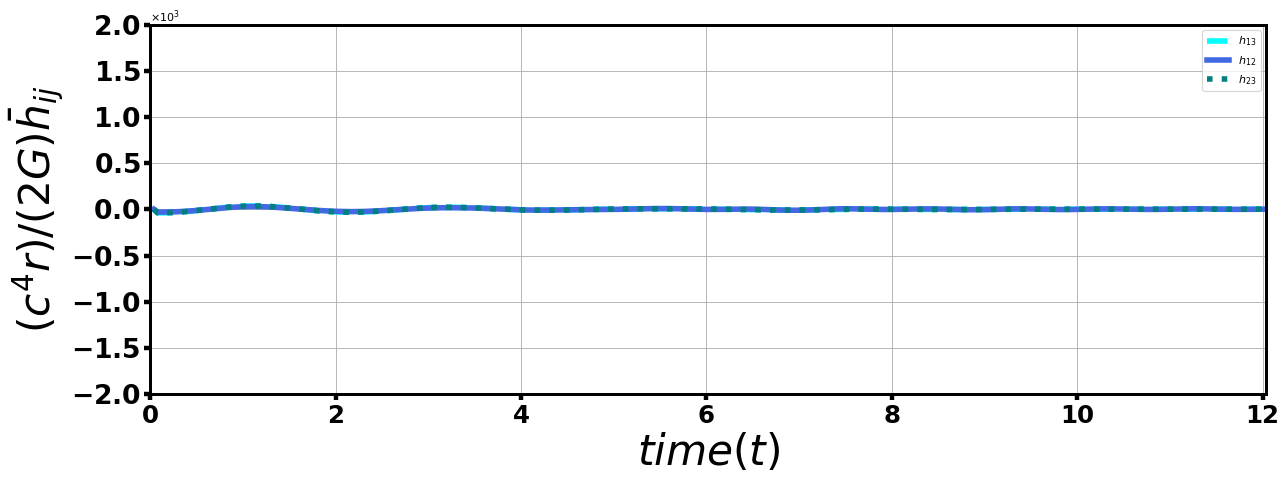}\\
\includegraphics[width=.90\columnwidth]{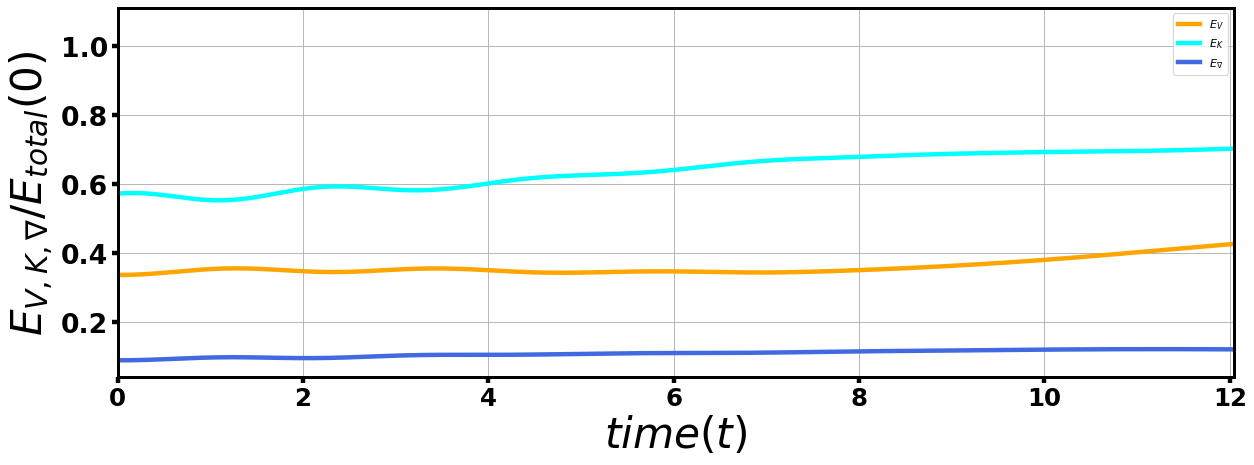}\\
\includegraphics[width=.90\columnwidth]{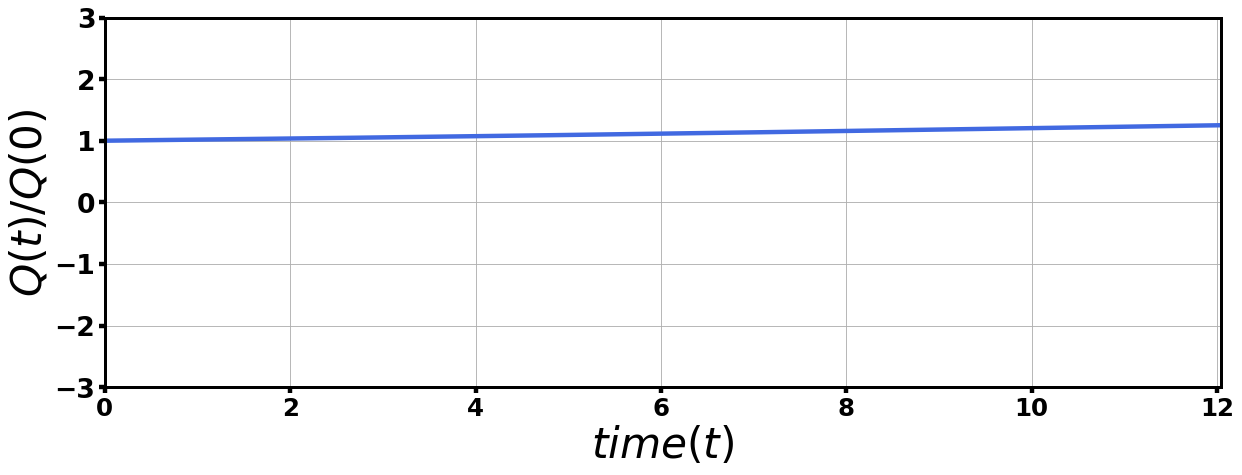}
\caption{Q-ball/Q-ball collision ($\omega^2=0.847^2$) at $t\simeq 6$. High velocity $v=0.6c$ pass through event with minimal interactions as in Figure 3.}
\label{fig:figure10}
\end{figure}

\begin{figure}[H]
\centering
\includegraphics[width=.90\columnwidth]{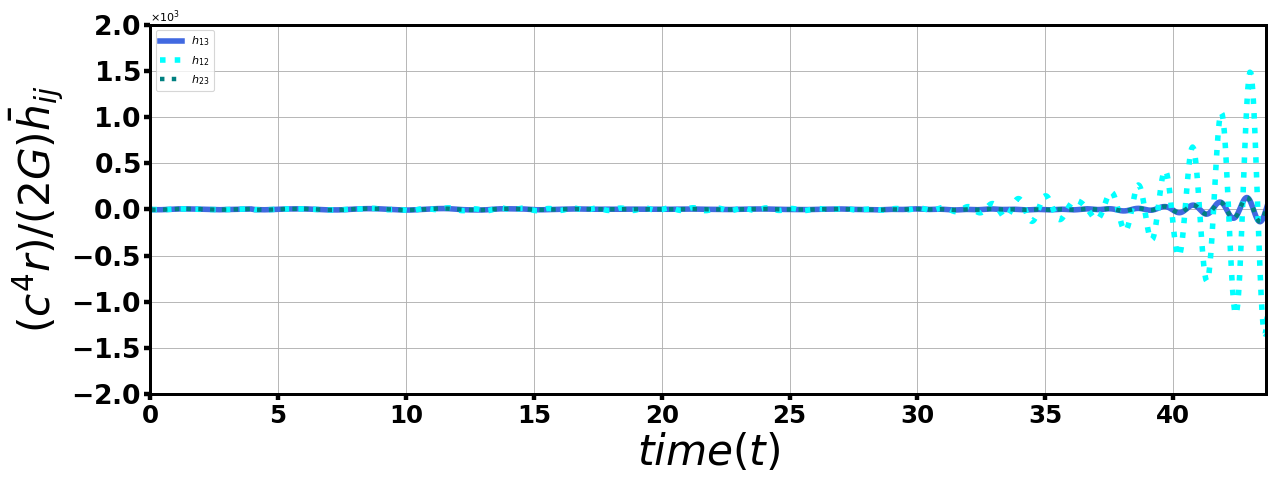}\\
\includegraphics[width=.90\columnwidth]{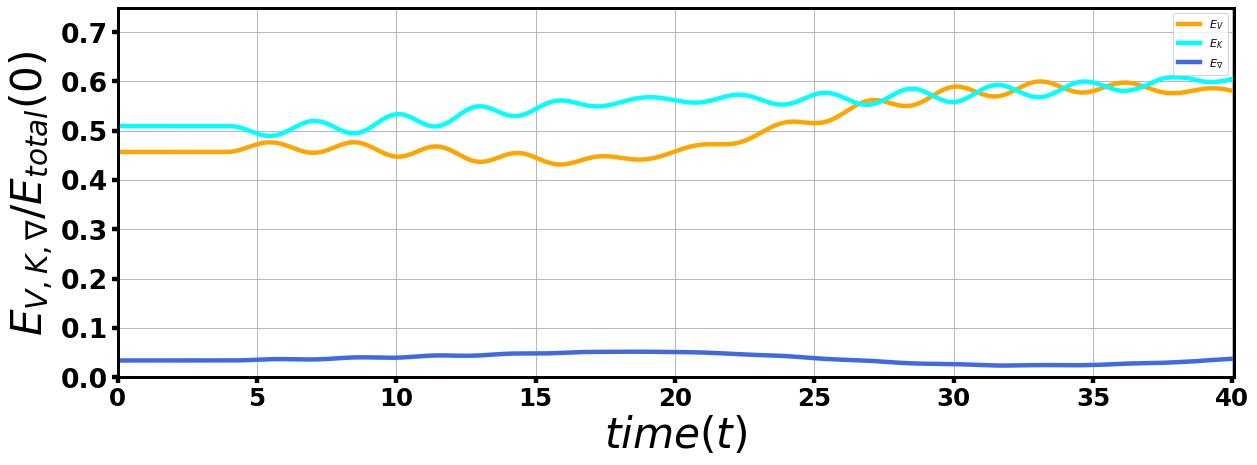}\\
\includegraphics[width=.90\columnwidth]{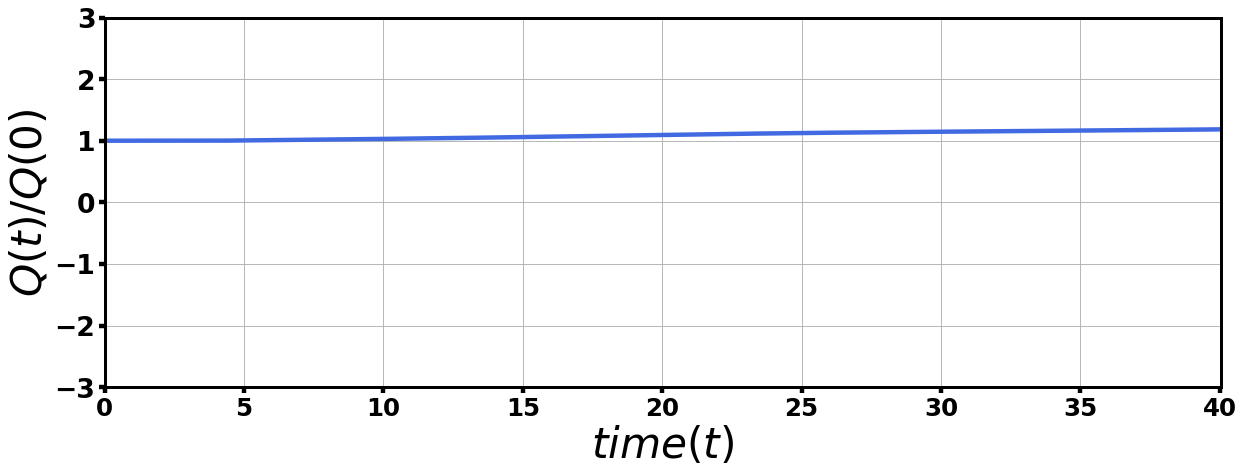}
\caption{
    Q-ball/Q-ball collision ($\omega^2=0.806^2$, $v=0.14c$) at $t\simeq5$ forming radial oscillation cycle to $t \simeq 40$ as in Figure 4.}
\label{fig:figure11}
\end{figure}

\begin{figure}[H]
\centering
\includegraphics[width=.90\columnwidth]{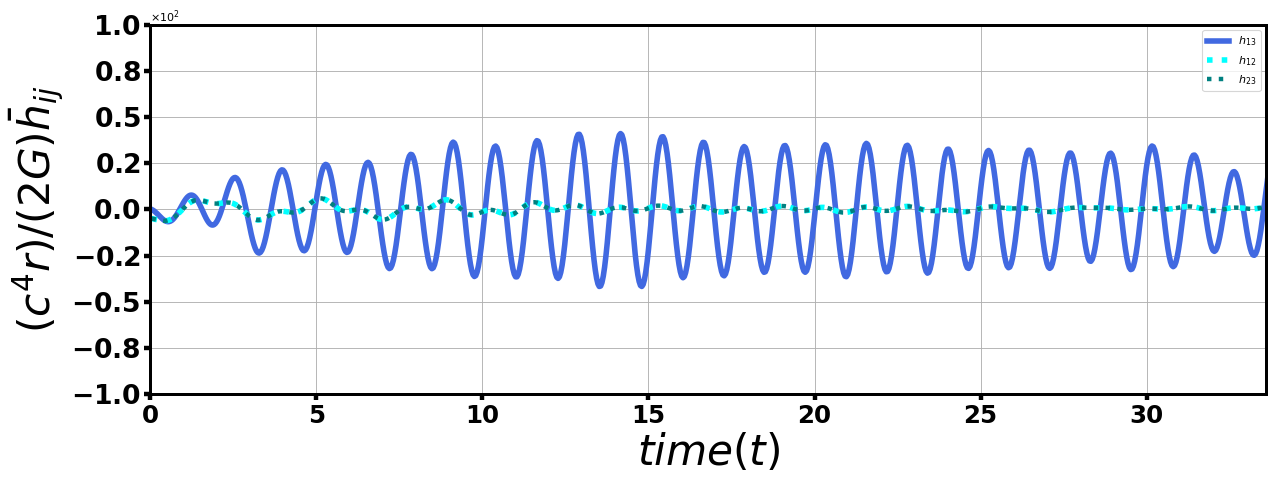}\\
\includegraphics[width=.90\columnwidth]{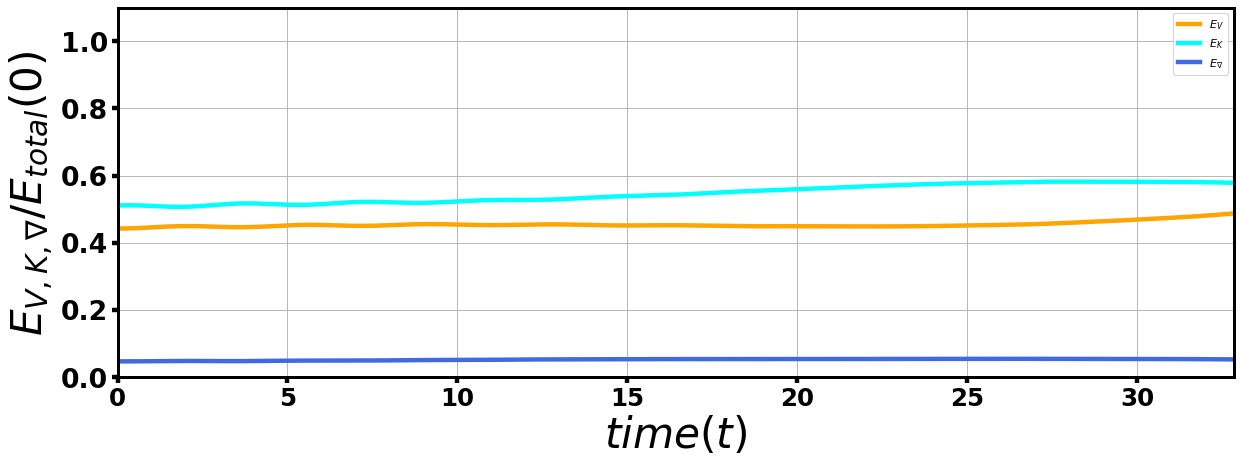}\\
\includegraphics[width=.90\columnwidth]{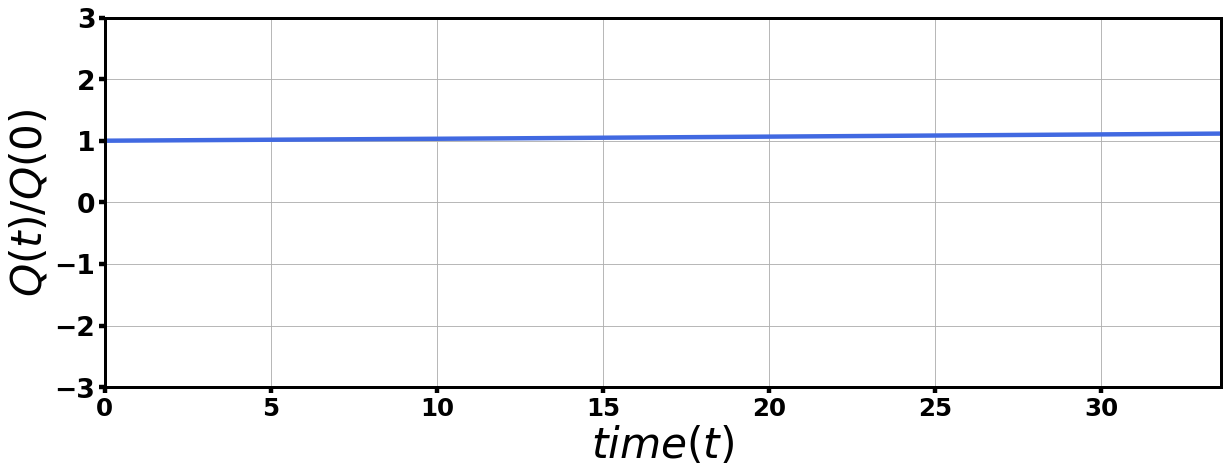}
\caption{Q-ball/Q-ball collision ($\omega^2=0.806^2$, $v=0.2c$) with central ring and two Q-balls ejected as in Figure 5.}
\label{fig:figure12}
\end{figure}
\begin{figure}[H]
\centering
\includegraphics[width=.900\columnwidth]{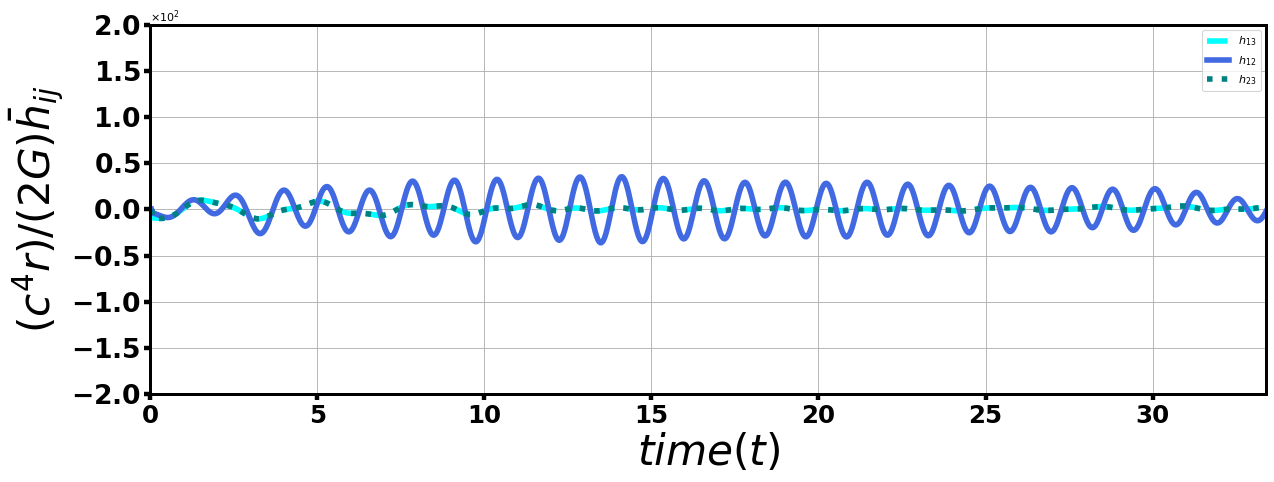}\\
\includegraphics[width=.90\columnwidth]{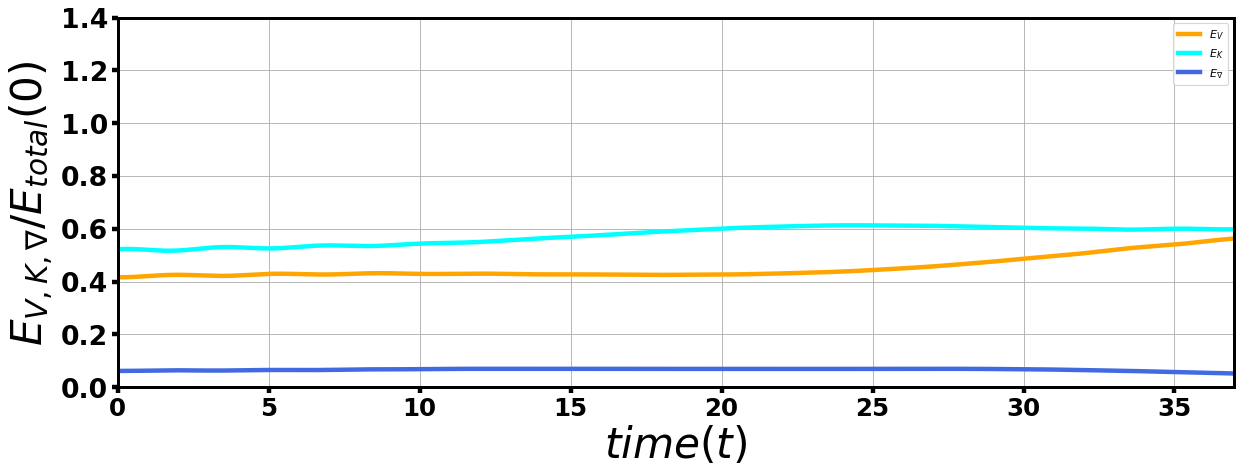}\\
\includegraphics[width=.90\columnwidth]{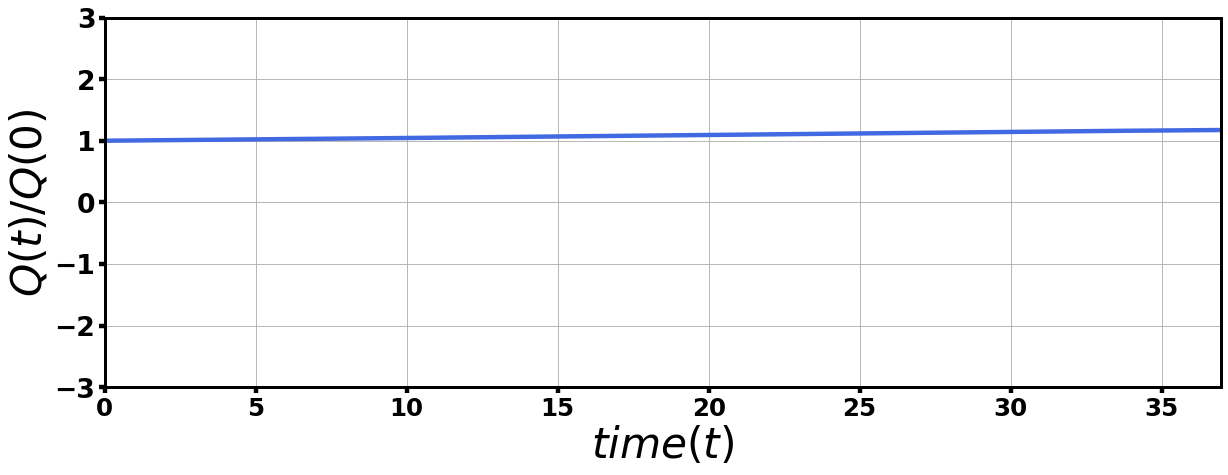}
\caption{Q-ball/Q-ball collision ($\omega^2=0.806^2$, $v=0.4c$) at $t\simeq0$ with ring ejection as in Figure 6.}
\label{fig:figure13}
\end{figure}
\begin{figure*}[th]
\centering
\includegraphics[trim={0.0 0.2cm 0.0 0.45cm},width=.33\textwidth, clip]{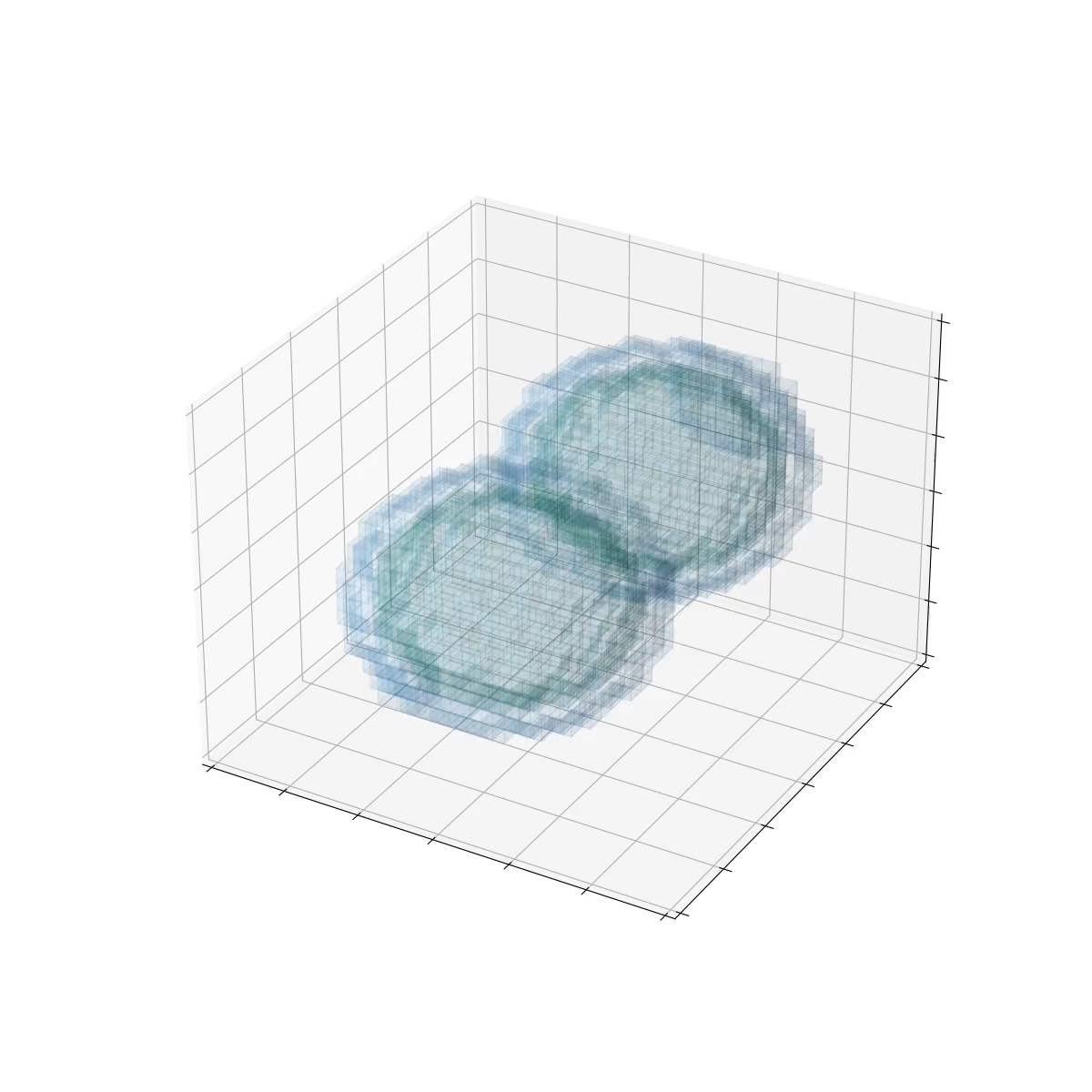}\hfill
\includegraphics[trim={0.0 0.2cm 0.0 0.45cm},width=.33\textwidth, clip]{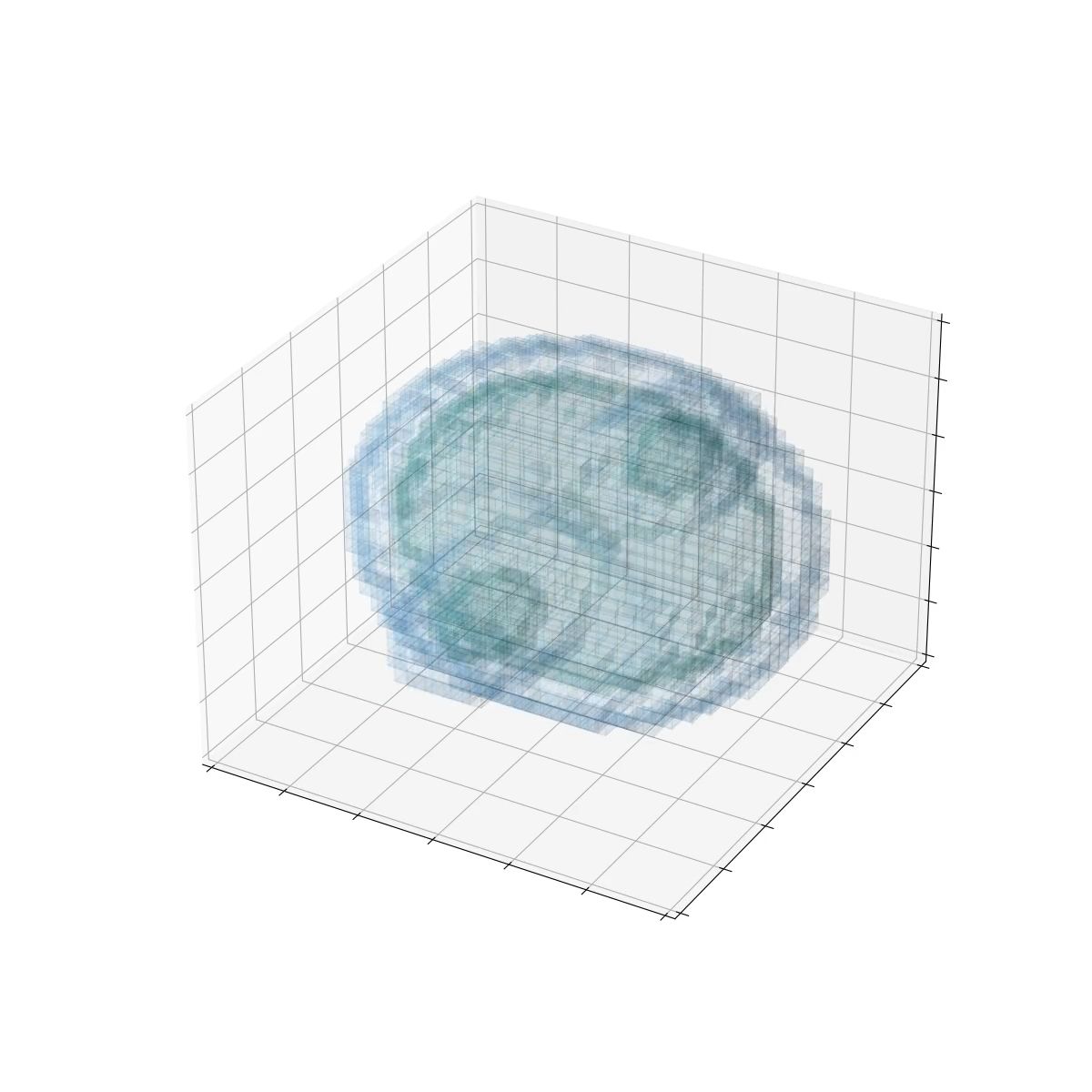}\hfill
\includegraphics[trim={0.0 0.2cm 0.0 0.45cm},width=.33\textwidth, clip]{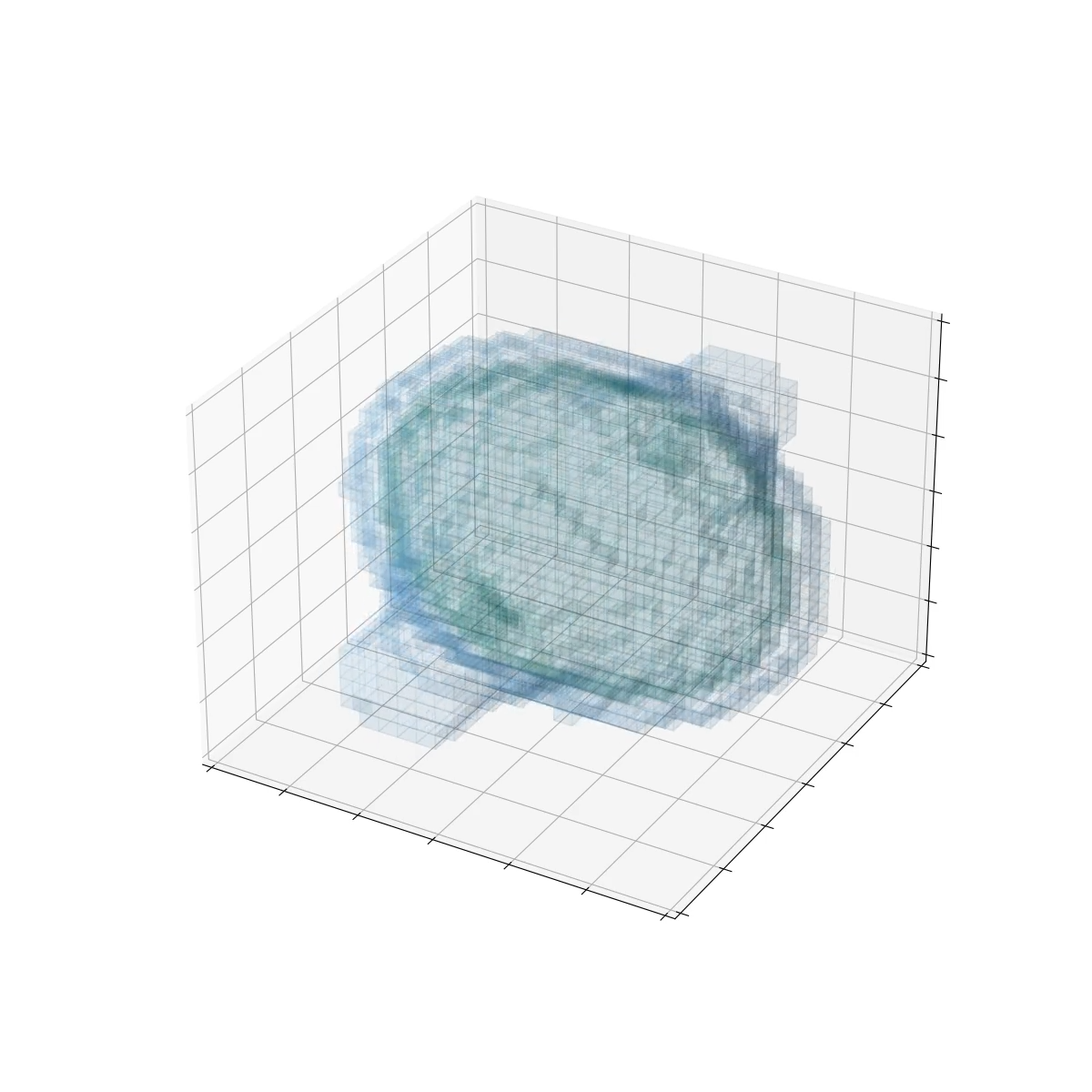}\\
\includegraphics[trim={0.0 0.2cm 0.0 0.45cm},width=.33\textwidth, clip]{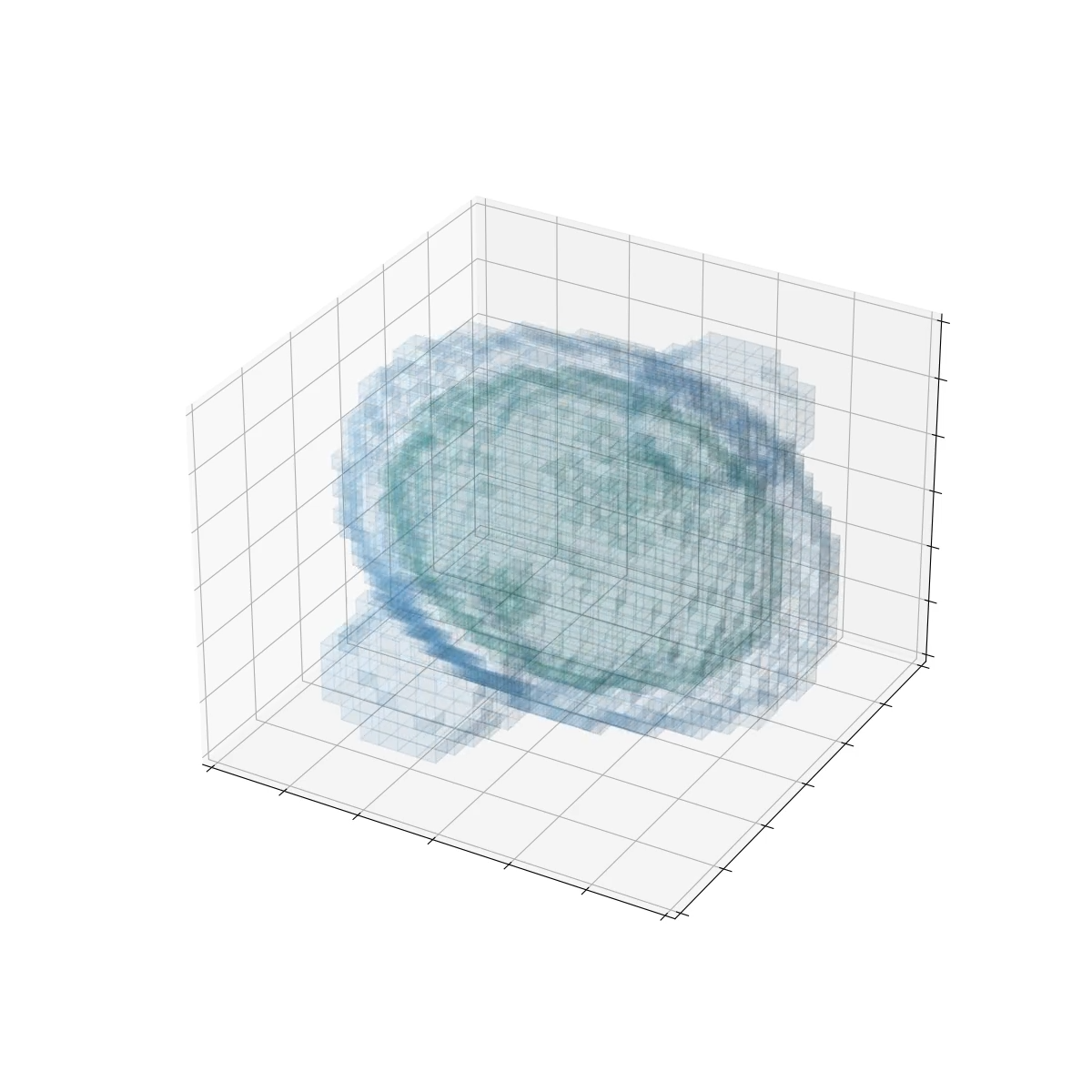}\hfill
\includegraphics[trim={0.0 0.2cm 0.0 0.45cm},width=.33\textwidth, clip]{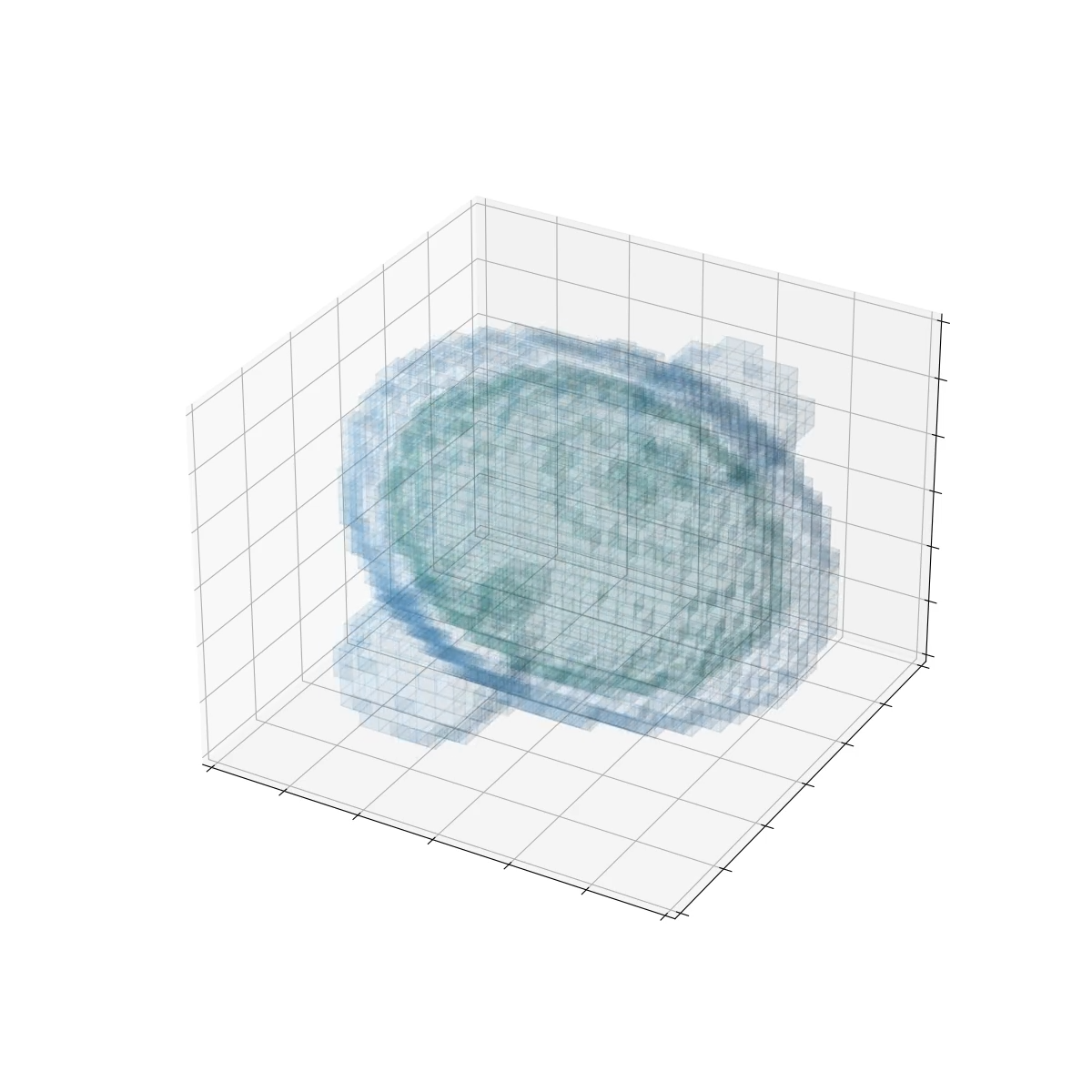}\hfill
\includegraphics[trim={0.0 0.2cm 0.0 0.45cm},width=.33\textwidth, clip]{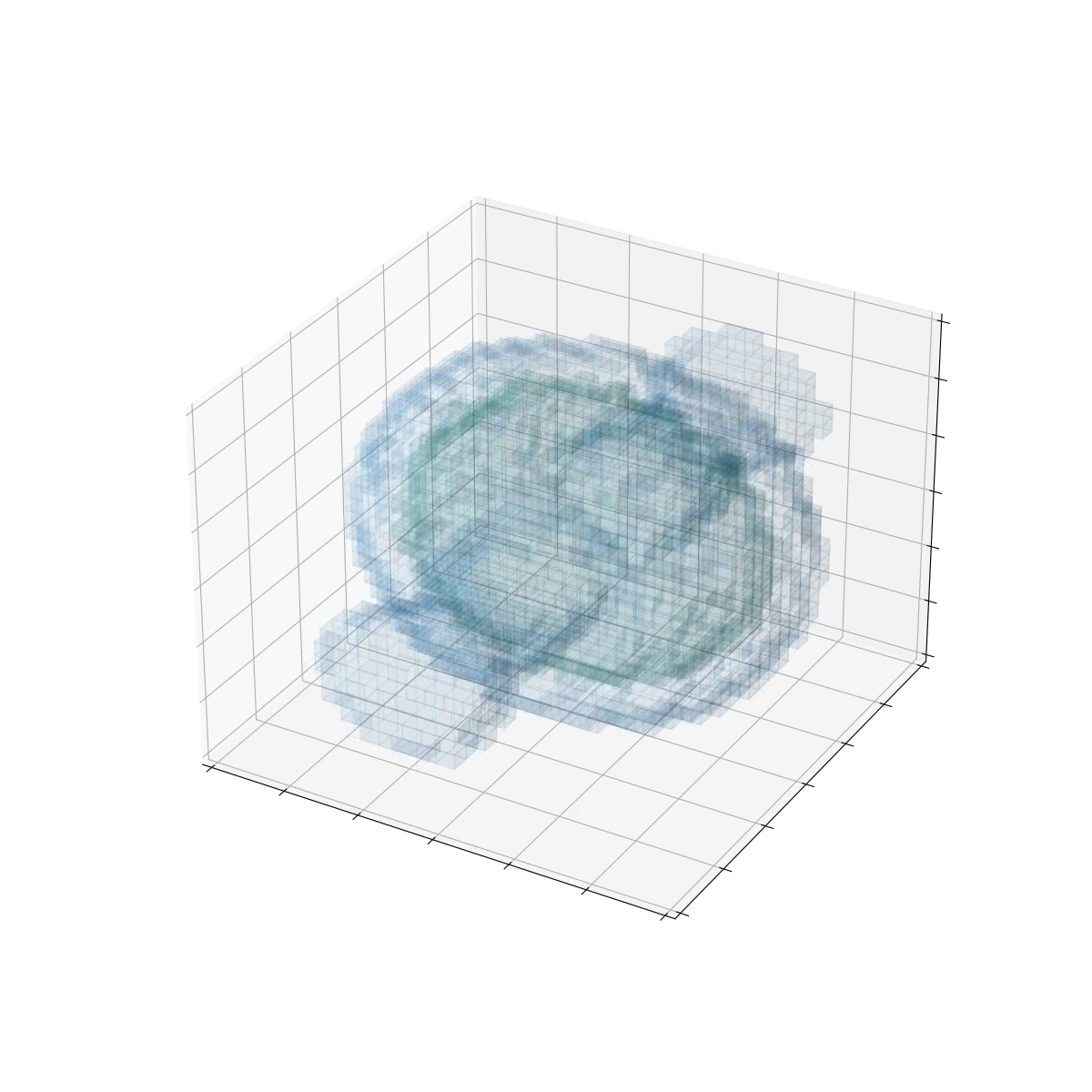}
\caption{The same collision as Fig. 5  with metric-perturbation conjugate field $\Pi_{\beta_{22}}$ in blue above scalar field energy density in green. The momentum conserving shift from mass largely distributed along the axis of collision, to the x-z symmetric ring  produces a novel gravitational wave signature in comparison to non-critical velocity collisions dominated by the scalar field energy density spike at the centre-of-mass.}
\label{fig:figure14}
\end{figure*}

\begin{figure}[b!]
\centering
\includegraphics[trim={0.0 0.2cm 0.0 0.25cm},width=.850\columnwidth, clip]{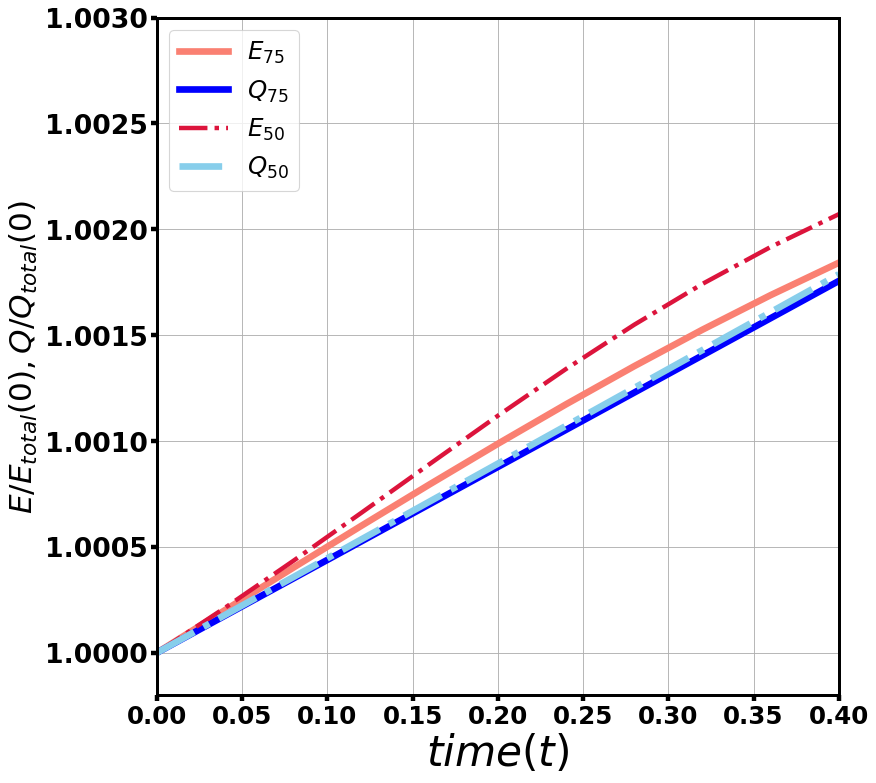}\\
\caption{Variation of energy and charge conservation for a single Q-ball in motion with lattice size $L=50$ in broken lines, while $L=75$ in solid lines. The time step size in both cases is $dt=0.02$.}
\label{fig:figure15}
\end{figure}

\bibliography{BIB}


\end{document}